\documentclass[twocolumn]{aastex631}

\newcommand{\lya}{Ly$\alpha$\ }

\newcommand{\angs}{\, {\rm \AA}}

\newcommand{\no}[1]{}

\def\lsim{~\rlap{$<$}{\lower 1.0ex\hbox{$\sim$}}}

\def\gsim{~\rlap{$>$}{\lower 1.0ex\hbox{$\sim$}}}

\usepackage{xcolor}

\usepackage{listings}
\usepackage{soul}

\usepackage[makeroom]{cancel}
\usepackage{verbatim}
\usepackage{amsmath}

\def\aap{A\&A}

\def\apjs{ApJS}
\def\apjl{ApJL}

\shorttitle{LIMFAST. I. A Semi-Numerical Tool for LIM}
\shortauthors{Mas-Ribas et al.}
\graphicspath{{./}{figures/}}

\submitjournal{ApJ}

\begin{document}

\defcitealias{Sun_2022P2}{Paper~II}

\title{\large{LIMFAST. I. A Semi-Numerical Tool for Line Intensity Mapping}}

\email{gsun@astro.caltech.edu}

\author{Llu\'is Mas-Ribas}
\affiliation{Jet Propulsion Laboratory, California Institute of Technology, 4800 Oak Grove Drive, 
Pasadena, CA 91109, USA}
\affiliation{California Institute of Technology, 1200 E. California Blvd, Pasadena, CA 91125, USA\\ }

\author{Guochao Sun}
\affiliation{California Institute of Technology, 1200 E. California Blvd, Pasadena, CA 91125, USA\\ }

\author{Tzu-Ching Chang}
\affiliation{Jet Propulsion Laboratory, California Institute of Technology, 4800 Oak Grove Drive, 
Pasadena, CA 91109, USA}
\affiliation{California Institute of Technology, 1200 E. California Blvd, Pasadena, CA 91125, USA\\ }

\author{Michael O. Gonzalez}
\affiliation{California Institute of Technology, 1200 E. California Blvd, Pasadena, CA 91125, USA\\ }

\author{Richard~H.~Mebane}
\affiliation{Department of Astronomy and Astrophysics, University of California, Santa Cruz, 1156 High Street, Santa Cruz, CA 95064, USA \\
\copyright 2023. All rights reserved.}


\begin{abstract}

We present LIMFAST, a semi-numerical code for simulating high-redshift galaxy formation and cosmic reionization as revealed by multi-tracer line intensity mapping (LIM) signals. LIMFAST builds upon and extends the 21cmFAST code widely used for 21 cm cosmology by implementing state-of-the-art models of galaxy formation and evolution. The metagalactic radiation background, including the production of various star-formation lines, together with the 21 cm line signal tracing the neutral intergalactic medium (IGM), are self-consistently described by photoionization modeling and stellar population synthesis coupled to the galaxy formation model. We introduce basic structure and functionalities of the code, and demonstrate its validity and capabilities by showing broad agreements between the predicted and observed evolution of cosmic star formation, IGM neutral fraction, and metal enrichment. We also present the LIM signals of 21 cm, Ly$\alpha$, H$\alpha$, H$\beta$, [\ion{O}{2}], and [\ion{O}{3}] lines simulated by LIMFAST, and compare them with results from the literature. We elaborate on how several major aspects of our modeling framework, including models of star formation, chemical enrichment, and photoionization, may impact different LIM observables and thus become testable once applied to observational data. LIMFAST aims at being an efficient and resourceful tool for intensity mapping studies in general, exploring a wide range of scenarios of galaxy evolution and reionization and frequencies over which useful cosmological signals can be measured. 

\end{abstract}

\section{Introduction} \label{sec:intro}

Line intensity mapping (LIM) provides a statistical approach to the study of the formation and evolution of galaxies and large-scale structure (LSS) in the universe. Compared to more traditional observational techniques that are limited to the individually detectable bright sources, LIM measurements of galaxies take into account the emission produced by the entire galaxy population present in large areas of the sky \citep{Madau1997,Suginohara1999,Visbal2010,kovetz2019}. This characteristic is especially relevant for high redshift studies, including the epochs of reionization and cosmic dawn. The LIM approach can prove beneficial here because the faint end of the galaxy population may have played a major role during these early times, but this is difficult to explore directly \citep[e.g.,][]{Fontanot2012,Choudhury2007,Robertson2015,Yue2018}. 

A number of emission lines resulting from different radiative processes and phases of the interstellar medium (ISM) are taken into account for LIM studies, the usual ones being the [\ion{C}{2}] line at 158\,$\mu$m \citep[e.g.,][]{Gong2012,Silva2015,Yue2015,Dumitru2019,YueFerrara2019,Sun2021}, those of the CO molecule \citep[e.g.,][]{Righi2008,Gong2011,Lidz2011,Pullen2013,Li2016,Chung2019,Ihle2019}, the hydrogen 21 cm spin-flip transition \citep[e.g.,][]{Scott1990,Madau1997,Furlanetto_2006PhR,Chang2008,Visbal2009,Chang2010,PritchardLoeb2012,Switzer2013,LiuShaw2020}, and the potentially bright rest-frame optical/ultraviolet (UV) lines such as H$\alpha$, H$\beta$, Ly$\alpha$, \ion{He}{2}, [\ion{O}{2}], and [\ion{O}{3}] among others \citep[e.g.,][]{Silva2013,Pullen2014,Visbal2015,Comaschi2016,Heneka2017,Gong2017,Visbal2018,Masribas2020,Heneka2021,Kannan_2022_LIM,Padmanaban2021,Parsons2022}. Distinct from past theoretical studies focusing on individual line tracers, an increasing number of recent modeling efforts aim at building a unified and self-consistent framework for multi-tracer investigations, which may provide a much more detailed and complete picture of high-redshift universe of interest \citep[e.g.,][]{Sun2019,Yang2021,Bethermin2022,Kannan_2022_LIM}. 

Numerically modeling LIM data is of major importance to guide future missions and experiments, but it is computationally challenging because the statistical power of LIM resides in the analysis of line emission signals over large areas of the sky and at multiple frequencies that are sensitive to small-scale physics. Simulations need to include both detailed processes related to star formation and the emission and transport of radiation in large cosmological volumes. This combination of a broad range of dynamical scales is demanding: numerical simulations accounting for resolved ISM-scale galaxy physics typically only exist for a small number of galaxies \citep[e.g.,][]{Hopkins2018,katz2019,Pallotini2019,kannan2020}. On the other hand, simulations covering large volumes tend to lack numerical precision at the smallest scales, and the processes connected to star formation are often modeled by means of sub-grid prescriptions \citep[e.g.,][and the review by \citealt{Vogelsberger2020}]{Vogelsberger2014,Eide2018,Eide2020,Shen2020,Shen2022,Kannan_2022_THESAN,Kannan_2022_LIM,Lewis2022}. Overall, in all cases these simulations typically require substantial computational resources and are thus not well-suited for parameter space exploration and model inference. 

With these limitations and constraints in mind, we present here LIMFAST, a semi-numerical tool designed for flexible modeling of high-redshift LIM signals. LIMFAST aims at self-consistently simulating line emission from galaxies and the intergalactic medium (IGM), over scales of several hundreds of Mpc and spanning the epoch of cosmic reionization, in a matter of hours with a current personal computer. LIMFAST builds upon and uses the 21cmFAST code \citep{Mesinger2007,Mesinger2011} to compute the underlying large-scale structure in large volumes of the universe. This computational step is rapidly achieved because 21cmFAST uses perturbation theory and analytical approaches to approximate the evolution of the density and velocity fields, as well as the formation of collapsed objects. LIMFAST inherits these calculations and applies analytic galaxy formation models adopted from \cite{Furlanetto_2021} to it to compute the radiation fields for a number of emission lines, in addition to the original 21 cm line from 21cmFAST. The progress of reionization is simultaneously computed also following the approach in 21cmFAST, but with extensions that self-consistently include the emission from galaxy populations that co-evolve with redshift. 

This paper is the first of a series that introduces the main structure of the LIMFAST code, and demonstrates the validity and capabilities of it to be used to address science questions related to the epoch of cosmic reionization. In the second paper, \citet[][Paper~II hereafter]{Sun_2022P2} presents the computations to include the [\ion{C}{2}] 158 $\mu$m and CO line emission and explores the effects and observational implications of different feedback and star formation prescriptions beyond the fiducial case presented in this work. Furthermore, a progenitor version of LIMFAST was presented and used in \cite{Parsons2022} to address the application of LIM to measure the average \ion{He}{2}/H$\alpha$ line ratio for inferring the initial mass function (IMF) of Population III (Pop~III) stars. The structure of this present paper is as follows: the models and key calculations implemented in LIMFAST are detailed in Section \ref{sec:code}, and the results from our fiducial simulation runs are shown and compared with the literature in Section \ref{sec:results}. We discuss some key considerations about these results, focusing on assumptions that give rise to qualitative differences in observables that may be tested by future LIM data in Section \ref{sec:discussion}, before concluding in Section \ref{sec:conclusion}. A flat, $\Lambda$CDM cosmology consistent with recent measurements by Planck \citep{Planck_XIII} is assumed throughout. 
  
\section{LIMFAST: the Code}\label{sec:code} 

We detail the LIMFAST code below, after a brief description of 21cmFAST. In Section \ref{sec:sfxs}, we present the galaxy model by \cite{Furlanetto_2021} and its implementation in LIMFAST. We next describe in Section \ref{sec:sed} the usage of BPASS to create the stellar spectra, and the photoionization calculations with \textsc{cloudy} to obtain the nebular radiation. The details of the ionization computations in the IGM are presented in Section \ref{sec:ion_calc}, and the calculations of the intergalactic and background Ly$\alpha$, as well as the 21 cm emission, are presented in Sections \ref{sec:lyaigm}, \ref{sec:lyab} and \ref{sec:21cm}, respectively. Finally, in Section \ref{sec:rsd}, we describe the inclusion of redshift-space distortions (RSD) in the calculations.

LIMFAST builds upon and extends 21cmFAST \citep{Mesinger2007,Mesinger2011} after inheriting the large-scale density and velocity fields computed by the latter at each simulated cell. In detail, 21cmFAST computes evolving density and velocity 
fields from a set of initial conditions and Lagrangian perturbation theory \citep{Zeldovich1970,Scoccimarro1998}. Then, the code makes use of the extended Press-Schechter formalism \citep{Lacey1993,Somerville1999} to obtain the collapsed mass field. As will be described in the next section, LIMFAST makes use of the Sheth-Tormen halo mass function (HMF) to connect and derive the properties of galaxies from halos at each cell, following the prescriptions in \cite{Furlanetto_2021}. We stress that in 21cmFAST, this approach corresponds to the ``matter density field'' case, where individual halos are not resolved nor identified in the simulation; at each cell, the halo mass distribution follows the Sheth-Tormen HMF, and the total number of halos depends on the matter overdensity and the collapsed fraction in that region. 

After the collapsed mass field is obtained, 21cmFAST computes the ionization state of the IGM by comparing the cumulative number of ionizing  photons from sources and the number of neutral hydrogen atoms and recombinations in spherical regions, from large to small volumes. This method, first proposed by \cite{Furlanetto2004}, is analogous to the excursion-set formalism and it allows for scenarios of inhomogeneous reionization. Finally, the code computes the X-ray and Ly$\alpha$ radiation backgrounds and uses them to derive the spin and brightness temperature of the 21 cm emission. In the following sections, we will discuss extensions and variations of these calculations in LIMFAST, and refer the interested reader to \citet{Mesinger2007} and \citet{Mesinger2011} for more details on 21cmFAST. 

\begin{figure*}
 \centering
 \includegraphics[width=0.9\textwidth]{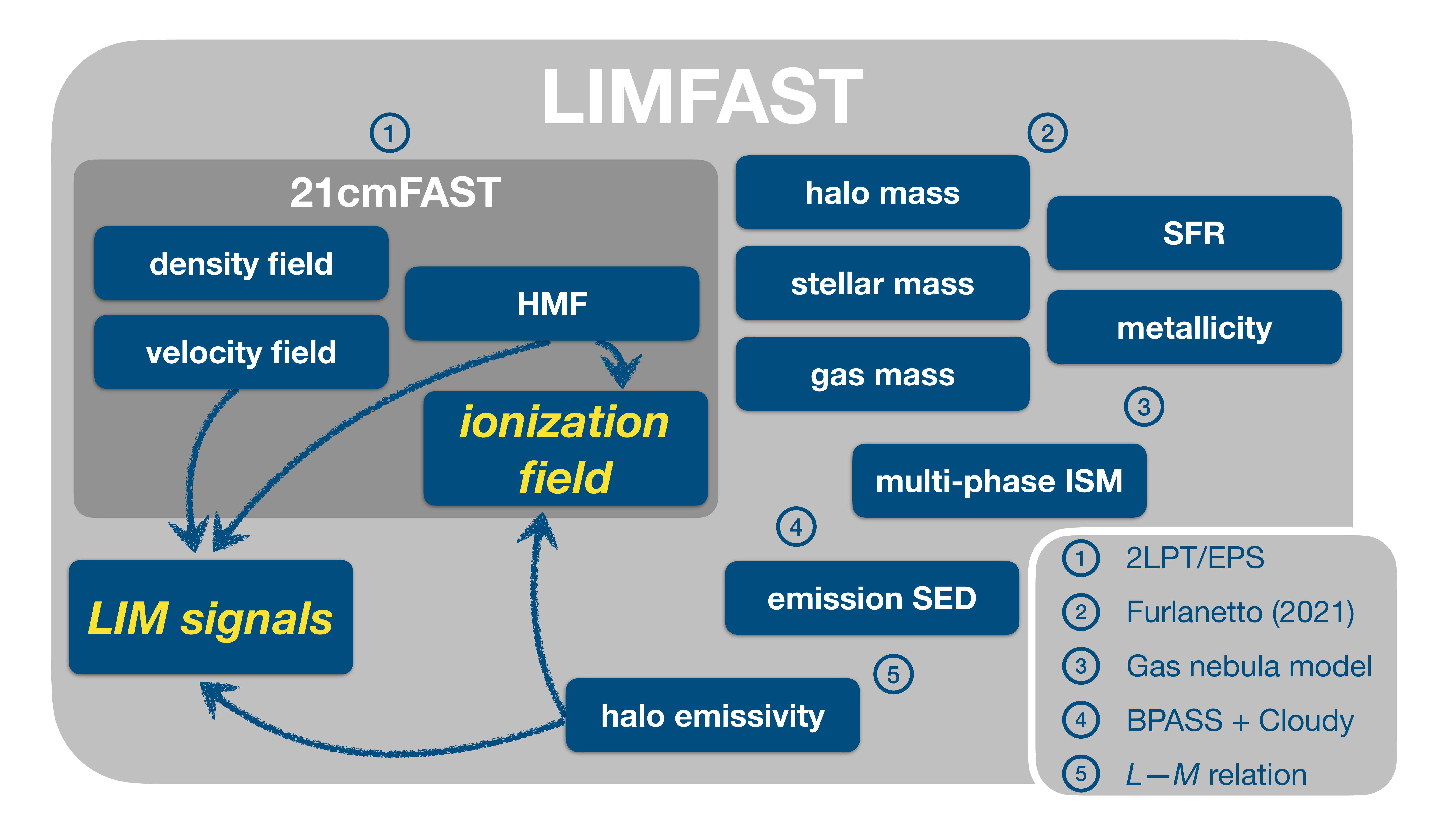}
 \caption{The basic, modular structure of LIMFAST. Inheriting the baseline method in 21cmFAST for approximating the formation of LSS and the partitioning of mass into dark matter halos, LIMFAST starts from assigning the SFR, as well as stellar, gas and metal masses as a function of halo mass and redshift using the \cite{Furlanetto_2021} galaxy models. These galaxy properties are combined with tabulated stellar and nebular SEDs and luminosities computed with BPASS and \textsc{cloudy}, and then yield the amount of line and ionizing continuum emission from the halos. The halo luminosities are integrated over the HMF computed from the density field (and offset by the velocity field in redshift-space calculations) for individual cells to generate the line intensity fields of interest, whereas the ionizing radiation is used to compute the ionization of the IGM as done in 21cmFAST.}
 \label{fig:flowchart}
\end{figure*}

Figure \ref{fig:flowchart} summarizes the basic ingredients of the LIMFAST code and its connection to 21cmFAST. Starting from the stellar, gas, and metal masses of a given halo derived from the \cite{Furlanetto_2021} galaxy models, the halo star formation rate and the metallicity are derived. These last two parameters, combined with the tabulated stellar and nebular spectra computed with BPASS and \textsc{cloudy}, then yield  the amount of line and ionizing emission from the halo. The halo luminosities in a given cell are integrated over the (conditional) HMF computed by 21cmFAST from the density field to generate the intensity field, with the RSDs being accounted for by the evolved velocity field. Meanwhile, the ionizing radiation is used to calculate the ionization state of the IGM.

\subsection{Galaxy Formation and Evolution Model}\label{sec:sfxs} 

We summarize next how the properties of galaxy-hosting halos are derived from the galaxy formation model introduced by \citet{Furlanetto_2021}. In \citet{Furlanetto_2021}, galaxies evolve due to the interplay between star formation and feedback. Star formation is fueled by a smooth accretion of mass onto the dark matter halo, and it is at the same time regulated by the feedback that ejects gas outside the galaxy via outflows. In this scenario, the co-evolution of gas, stellar, and metal masses can be described by 
\begin{equation}\label{eq:mgeq}
\dot{M}_{\rm g} = f_\mathrm{b}\dot{M} - (\mathcal{R} + \eta)\dot{M}_*~,
\end{equation}
\begin{equation}\label{eq:sfreq}
\dot{M}_* = M_{\rm g} / t_{\rm sf}~,
\end{equation}
and
\begin{equation}\label{eq:mzeq}
\dot{M}_Z = -(1+\eta)Z\dot{M}_* + y_Z \dot{M}_*~. 
\end{equation}
In Equation~(\ref{eq:mgeq}) above, $f_\mathrm{b} = \Omega_\mathrm{b}/\Omega_\mathrm{m} \approx 1/6$ is the baryon fraction, $\dot M$ is the mass accretion rate of the halo, $\mathcal{R} \approx 0.25$ denotes the fraction of mass available for star formation that resides in stars, and $\eta \gg 1$ accounts for the amount of mass ejected out of the galaxy by stellar feedback. In Equation~(\ref{eq:sfreq}), $t_{\rm sf} = t_{\rm orb}/\epsilon$ denotes the star-formation timescale, where $t_{\rm orb} \sim 18 \left[7/(1 + z)\right]^{3/2}$ Myr is the orbital timescale, and $\epsilon$ characterizes the temporal star formation efficiency per orbital timescale. In Equation~(\ref{eq:mzeq}), $Z \equiv M_Z/M_{\rm g}$ denotes the metallicity, and the metal yield factor $y_Z = 0.03$ \citep{Benson_2010PhR} is the fraction of stellar mass returned to the ISM in the form of metals. 
 
Now writing $M_{\rm g}$ as a fraction of the total accreted mass $M_\mathrm{a}=f_\mathrm{b} M$, one can define $X_{\rm g} \equiv M_{\rm g} / M_\mathrm{a}$ as the gas retention factor \citep{DM_2014MNRAS}, and similarly $X_* \equiv M_*/M_\mathrm{a}$. The previous equations can then be rewritten in terms of the dimensionless quantity $\tilde{M} \equiv M/M_0$, where $M_0$ denotes the halo mass at some initial redshift $z_0$, and taking derivatives with respect to redshift instead of time, namely $\tilde{M}^\prime = d \tilde{M} / d z$. With these substitutions, the system of differential equations describing the halo, gas, and stellar mass evolution with redshift finally equates
\begin{equation}
\frac{\tilde{M}^\prime}{\tilde{M}} = -|\tilde{M}_0^\prime|~,
\label{eq:mh_prime}
\end{equation}

\begin{equation}
\frac{\tilde{M}_{\rm g}^\prime}{\tilde{M}_{\rm g}} = -|\tilde{M}_0^\prime|  \left[ X_{\rm g}^{-1} - \frac{\epsilon(\mathcal{R}+\eta)}{|\tilde{M}_0^\prime| C_\mathrm{orb}} \left(\frac{1+z_0}{1+z}\right) \right]~,
\end{equation}

\begin{equation}
\tilde{M}_*^\prime = -\mathcal{R} \tilde{M}_{\rm g} \left[ \frac{\epsilon}{C_\mathrm{orb}} \left(\frac{1+z_0}{1+z}\right) \right]~,
\label{eq:ms_prime}
\end{equation}

\begin{equation}
\frac{\tilde{M}_Z^\prime}{\tilde{M}_Z} = \mathcal{R} \left(-1 - \eta + y_Z Z^{-1} \right) \left[ \frac{\epsilon}{C_\mathrm{orb}} \left(\frac{1+z_0}{1+z}\right) \right]~. 
\label{eq:ms}
\end{equation}
Here, $C_{\rm orb}=(1+z_0) t_{\rm orb} H(z)$ characterizes the parameter dependence of the orbital timescale, and $H(z)$ is the Hubble parameter at redshift $z$. 

In order to solve the above equations, the parameters denoting the feedback model, $\eta$ and 
$\epsilon$, also need to be specified. For the fiducial LIMFAST case introduced here, we adopt the momentum-driven feedback model by \citet{Furlanetto2017}, and present detailed dependencies on feedback prescriptions, as well as other star formation recipes, in \citetalias{Sun_2022P2}. In the momentum-driven feedback case, we set $\epsilon = 0.015$, and compute the term denoting the relation between the rate of gas expelled from the galaxy and the star formation rate as 
\begin{equation}
   \eta(M,z)= C \left( \frac{10^{11.5}}{M}\right)^{\xi} \left( \frac{9}{1+z}\right)^{\sigma}  ~,
\end{equation}
where we have adopted the parameter values $C=5$, $\xi=1/3$, and $\sigma=1/2$, consistent qualitatively with findings from abundance matching to the observed galaxy UV luminosity functions (UVLF) at $z\gtrsim5$ \cite[e.g.,][]{Mason2015, Sun2016, Mirocha_2017}. 

For the implementation of this galaxy model in LIMFAST, we have solved the above equations considering an initial redshift of $z=30$ and tabulated the results as a function of halo mass and redshift. These tables cover the redshift range between $z=5$ and $z=30$ in steps of $\Delta z =0.1$, and the halo masses range from $M=10^7\, M_\odot$ to $M=10^{16}\, M_\odot$ in 900 evenly-distributed logarithmic bins. LIMFAST then interpolates the tables to obtain the result for any combination of mass and redshift within these ranges. 

Once the above quantities are calculated, we can derive two other important observables, the star formation rate density (SFRD) and the mass-averaged metallicity. The SFRD in a simulation cell is obtained by integrating the star formation rate per halo over the (conditional) HMF of the cell at position $\boldsymbol{x}$ as 
\begin{equation}
 \dot{\rho}_*(\boldsymbol{x}, z)  =  \int {d}n/{d}M(\boldsymbol{x}, M, z) \,\dot{M}_*(M, z) \,{d} M ~,
\label{eq:sfrd}
\end{equation}
where ${\rm d}n/{\rm d}M$ denotes the HMF in 21cmFAST, consistent with the Sheth-Tormen \citep{Sheth1999} formalism with the correction by \cite{Jenkins2001}. For an individual halo, the metallicity 
at a given redshift is simply defined as $Z(M,z) \equiv M_Z(M, z)/ M_{\rm g}(M, z)$, whereas for a cell the metallicity is computed as the ratio of total metal mass to total gas mass in such a cell, namely 
\begin{equation}\label{eq:metal}
Z(\boldsymbol{x},z)  =  \frac{\int {d}n/{d}M(\boldsymbol{x}, M, z)\, M_Z(M, z) \,{d} M}{\int {d}n/{d}M(\boldsymbol{x}, M, z) \,M_{\rm g}(M, z) \,{d} M} ~.
\end{equation}
This calculation differs from that of the star formation rate because the metallicity is not an additive quantity and, therefore, one cannot integrate the metallicity per halo over the HMF. However, one may want to perform the metallicity per halo integration, and then divide it by the total number density of halos in the respective cell, to obtain the average metallicity per halo in such a volume. The integrals in the above two equations are performed from a minimum halo mass, $M_\mathrm{min}$, set to the atomic cooling halo mass at a virial temperature of $T_{\rm vir}=10^4\,$K, to a maximum halo mass of $M_\mathrm{max}=10^{16}\,M_{\odot}$. The value of $M_\mathrm{min}$ varies with redshift from $\sim 10^7\,M_\odot$ at $z=20$ to $\sim 10^8\, M_\odot$ at $z=5$. Considering the atomic cooling threshold here is valid because we do not account for Pop~III star formation in this work; one may want to include smaller halo masses corresponding to molecular cooling when accounting for that stellar population \citep[see, e.g.,][and references therein for further discussions]{Mebane2018,Munoz2022,Parsons2022}. Meanwhile, it should also be considered as an optimistic case for the formation of Pop~II stars, since their formation in low-mass (but still above this limit) systems may be regulated by radiative feedback due to reionization \citep{Yue2014,Yue2018}, although the most recent constraints on the UVLF faint-end slope from deep lensed galaxies have yet to show the presence of a turnover \citep{Bouwens_2022}. Moreover, while so far only kinetic feedback from supernova explosions are considered in LIMFAST, we note that reionization feedback can also leave detectable imprints on intensity mapping signals \cite[see e.g.,][]{Mirocha2022}. 

Finally, we define the comoving luminosity density from halos in a cell as 
\begin{equation}
 \rho_l(\boldsymbol{x}, z) = \int {d}n/{d}M(\boldsymbol{x}, M, z)\, l(M, z)\, {d} M~,
\label{eq:ldens}
\end{equation}
where $l(M, z)$ is the luminosity of a halo at a given redshift derived from the SFR of the halo and the line luminosity per unit SFR (to be detailed in the next section), and the integration limits are the same as before\footnote{We do not account for scatter in the relation between luminosity and halo mass in this work. This could be incorporated by considering a distribution of luminosity values instead of a single value in Equation~(\ref{eq:ldens}) and a random sampling of halos from the HMF.}. Then, the observed specific intensity is derived from the luminosity density as 
\begin{equation}\label{eq:intens}
    I_{\nu}(\boldsymbol{x}, z) =  \frac{c}{4\pi} \frac{\rho_{l}(\boldsymbol{x}, z)}{\nu_0\,H(z)}~,
\end{equation}
where $c$ is the speed of light and $\nu_0$ is the rest-frame frequency of the emission line of interest. In Section~\ref{sec:results} and beyond, we will refer to values of these quantities averaged over the entire co-eval simulation box in order to assess their redshift evolution, e.g., $\bar{I}_{\nu}(z) \equiv N^{-1}_\mathrm{cell} \sum_{i=0}^{N_\mathrm{cell}} I_{\nu}(\boldsymbol{x_{i}}, z)$.

\subsection{Stellar and Nebular SEDs}\label{sec:sed}

In this work, we account only for normal, Population II (Pop~II) stellar populations, and leave the inclusion of Pop~III stars to future implementations \citep[see, e.g.,][for previous work on implementations of Pop~III stars in 21cmFAST]{Mebane2018, Qin2020, Qin2021,Tanaka2021,Munoz2022,Parsons2022}. 

The original 21cmFAST code considers one single, redshift-independent stellar spectral energy distribution (SED) describing the Pop II stellar population, and uses its properties for the calculations of the ionization of the IGM and the \lya radiation background. The SED used by 21cmFAST assumes a Scalo IMF \citep{Scalo1998}, a metallicity of 0.05$Z_\odot$, and continuous star formation for 100\,Myr \citep{Barkana2005}. LIMFAST replaces this SED by a set of 13 metallicity-dependent stellar SEDs, and uses them to compute the ionization state of the IGM, the Ly$\alpha$ background, and the nebular line emission. This parameterization allows us to connect the processes of galaxy evolution and reionization consistently by using different SEDs that trace the redshift evolution of metallicity. 

The stellar SEDs used by LIMFAST are computed by using BPASS v2.1 \citep{Eldridge2017}, assuming a single-star and constant star formation mode with an age of 100\,Myr, and a Salpeter IMF \citep{Salpeter1955} with stellar masses within the range $0.5 – 100\, M_\odot$. The 13 SEDs differ from each other by their metallicity value, and they respectively account for the  default absolute BPASS metallicity values of $Z_\star=$ [$ 10^{-5}$, $10^{-4}$, $ 10^{-3}$, $ 2\times 10^{-3}$, $ 3\times 10^{-3}$, $ 4\times 10^{-3}$, $6\times 10^{-3}$, $8 \times 10^{-3}$, $ 10^{-2}$, $1.4 \times 10^{-2}$, $ 2\times 10^{-2}$, $ 3\times 10^{-2}$, $4 \times 10^{-2}$]. BPASS provides this range of metallicities that is well suited to reach the low metallicity values that may occur at the redshifts of reionization, as well as the possible 
higher values in massive objects. The number of ionizing photons per stellar baryon in these SEDs spans the range $N_\mathrm{ion} \sim 2500$--6000, where higher numbers are for more metal-poor SEDs. For comparison, the single value for the 
number of ionizing photons used in 21cmFAST is 4361. 

For the calculation of the nebular line emission, we use our stellar SEDs as the incident spectrum in the photoionization code \textsc{cloudy} \citep[version 17.02,][]{Ferland2017}, and the following quantities describing the nebular medium; we consider a gas density of $n_\mathrm{H} = 10^2$ $\mathrm{cm}^{-3}$ typical for \ion{H}{2} regions \citep{Byler2017}, a distance between the radiation source and the medium of $r = 10^{19}$ cm, and the ionization parameter values $\log U = [-4,\,-3.5,\,-3,\,-2.5,\,-2,\,-1.5,\,-1]$. 
Changing the value of $U$ with the other quantities fixed would imply that the 
number of ionizing photons also changes. In practice, however, the number of photons is fixed by the incident SED, so we renormalize the resulting nebular emission by the corresponding default number of photons in the SEDs in all cases. In other words, this approach would be equivalent to vary the value of $r$ or $n_{\rm H}$ to obtain different ionization parameter values 
while keeping a constant number of photons \citep[see similar procedures in][and \citealt{Xiao2018}]{Byler2017}. For each SED case, we assume the gas and the SED to have the same metallicity (i.e., equal stellar and gas-phase metallicities). We then perform photoionization calculations combining these metallicity and ionization parameter values and tabulate the emission results in units of luminosity per unit of star formation rate. For a given pair of metallicity and ionization parameter values describing a halo, LIMFAST then linearly interpolates the tabulated results to derive the luminosity per unit star formation rate in that halo, $l(M,z)$. In Appendix~\ref{sec:udependence}, we show the luminosity of star-formation lines of interest as a function of metallicity and ionization parameter as the key input to LIMFAST from the spectral synthesis and photoionization modeling using BPASS and \textsc{cloudy} (see Figure \ref{fig:udependence}). 

We do not consider dust in these sub-grid calculations and leave its implementation to future versions of the code, because the effect of dust attenuation remains highly uncertain for high-redshift galaxies \citep{Casey_2014,Casey_2018,Capak_2015,Popping_2017}. Not including dust implies that the default escape fraction of nebular radiation in  LIMFAST is 100\%, 
where we have presently also neglected the effect of neutral hydrogen on the escape and transfer of \lya emission. This is also the case for the damping wing absorption of \lya photons by intervening patches of neutral IGM along the line of sight (LOS). In other words, the line emission presented here is the intrinsic one. However, the calculation of the intrinsic 
emission allows the user to apply desired custom attenuation effects a posteriori both at a cell level and along any specific LOS, e.g., applying 
analytical extinction prescriptions that depend on redshift or other parameters \citep{Gong2017}, or treating the absorption and scattering of Ly$\alpha$ photons by the intergalactic \ion{H}{1} \citep{MF2008, Heneka2017}. Alternatively, one could implement the effects of 
dust directly on the SEDs (e.g., through a wavelength-dependent attenuation curve) resulting in attenuated 
nebular spectra \citep{Byler2017}, or by linking the dust properties 
to the halo metallicities resulting from the LIMFAST simulations.

Finally, the default calculations assume a constant escape fraction of ionizing photons of 10\% throughout. This value is chosen such that the resulting reionization history is broadly consistent with current 
constraints as further discussed in Section \ref{sec:results}. In practice, one can obtain different reionization histories in LIMFAST by changing the value 
of the escape fraction, as well as by varying the star formation parameters, such as feedback mode, star formation law, or $M_\mathrm{min}$ in the above galaxy model. We explore how these variations may impact the reionization history and morphology, as well as the corresponding multi-tracer LIM signals in \citetalias{Sun_2022P2}.

\subsection{Ionization Calculation}\label{sec:ion_calc}

As mentioned in the previous section,  21cmFAST uses a fixed value of 4361 ionizing photons per stellar baryon in the ionization calculations at all redshifts. Following our approach of metallicity-dependent SEDs, LIMFAST instead computes a number of ionizing photons that depends and evolves with metallicity. In detail, LIMFAST computes the number of ionizing photons for a halo of a given metallicity by linearly interpolating the number of ionizing photons from the two SEDs with metallicities closer to that of the halo. Therefore, because the metallicity of the halos changes with time, the number of ionizing photons also evolves with redshift. Finally, we have not considered radiation transfer effects that may produce further spatial variations in the ionization calculations (see, e.g., \citealt{DF_2022} for a recent discussion on the propagation of ionizing radiation and its effects on 21cmFAST, and \citealt{Lewis2022}).

\subsection{IGM Ly$\alpha$ Emission} \label{sec:lyaigm}

The IGM Ly$\alpha$ emission denotes here the Ly$\alpha$ radiation produced in situ in the IGM due to the recombination of ionized gas leading to the formation of \ion{H}{1}. We ignore collisional effects that may also lead to the production of Ly$\alpha$ because these are expected to be subdominant compared to recombination (see further discussions in \citealt{Silva2013} and \citealt{Comaschi2016}).

The comoving Ly$\alpha$ luminosity density from the recombination of IGM gas in a cell can be expressed as 
\begin{equation}
    \rho_{\rm Ly\alpha,\,rec}^{\rm IGM} (z) =  f_{\rm rec} \, E_{\rm Ly\alpha}\,\dot N_{\rm rec}(z)\,n_{\rm H \textsc{ii}}(z)~,
\end{equation}
where $f_{\rm rec} = 0.66$ denotes the fraction of recombinations producing  Ly$\alpha$ photons, $E_{\rm Ly\alpha} = 1.637 \times 10^{-11}\, {\rm erg}$ is the energy of the Ly$\alpha$ transition, and $\dot N_{\rm rec}(z)$ represents the recombination rate per baryon computed by the original 21cmFAST code. 
The last term above equates 
\begin{equation}
    n_{\rm H \textsc{ii}}(z) = [1- x_{\rm H \textsc{i}}(z)]\, n_{\rm b}\, [1 + \delta(z)]  ~,
\end{equation}
and it describes the comoving number density of ions in the IGM. Here, $x_{\rm H \textsc{i}}(z)$ denotes the neutral hydrogen fraction, $\delta(z)$ corresponds to the matter overdensity, and $n_{\rm b}$ is the present day comoving number density of baryons. The comoving luminosity density can be finally converted to the observed intensity by means of Equation (\ref{eq:intens}) as above.  

Although the IGM recombination process may produce diffuse emission of other hydrogen lines, we ignore their contribution. Besides Ly$\alpha$, H$\alpha$ would be the brightest of these hydrogen lines, but given its recombination coefficient and the transition probabilities  between the atomic energy level of the hydrogen atom, the expected luminosities for H$\alpha$ are expected to be around one order of magnitude fainter than those of Ly$\alpha$.

\subsection{Ly$\alpha$ Background}\label{sec:lyab}

Another component contributing to the Ly$\alpha$ radiation field is that produced by the scattering of high energy UV photons that, while traveling through the IGM, redshift into the frequency of the Lyman series lines in the rest frame of the IGM gas. When these initially high energy photons reach the frequencies of the Lyman lines, they are susceptible to be absorbed by the neutral \ion{H}{1} and they can subsequently lead to the processes of resonant scattering or a down-cascade by the electron in the atom that may ultimately produce Ly$\alpha$ radiation. 

We follow the calculation of the Ly$\alpha$ background in 21cmFAST, but we use our set of SEDs consistently instead of a single SED. In detail, each of the 13 stellar SEDs is tabulated to account for the number of photons in between the first 23 energy levels of the hydrogen atom as in 21cmFAST (see details of this calculation in \citealt{Barkana2005} and \citealt{Mesinger2011}). Then,  interpolation is used to find the photon number corresponding to metallicities within those of the two nearest SEDs as done for the line and ionizing emission.

With the Ly$\alpha$ background in the frame of the gas in units of photon rate per unit frequency, area and steradian, $J_\alpha$, the observed intensity can be obtained as \citep{Silva2013} 
\begin{equation}
    I^\mathrm{BG}_\mathrm{Ly\alpha} = \frac{6 \,E_{\rm Ly\alpha}}{(1+z)^4} J_\alpha ~,
\end{equation}
where the term in the denominator accounts for the redshift dimming of the surface brightness.

\subsection{21 cm Signal} \label{sec:21cm}

The 21 cm signal sourced by the neutral hydrogen in the IGM is often expressed as a 
differential brightness temperature that can take a positive or negative sign depending on whether the 
gas is emitting or absorbing 21 cm radiation. This differential brightness temperature can be written as  \citep{Furlanetto_2006PhR}
\begin{align}\label{eq:tbeq}
    \delta T_b (z) & \approx  27\, x_{\rm H \textsc{i}}(z)\, [1 + \delta(z)] \, {\left|1 + \frac{1+z}{ H(z)} \frac{dv_\parallel}{dr_\parallel}\right|}^{-1}  \nonumber \\
    & \times \left[ 1 - \frac{T_\gamma (z)}{T_S(z)} \right] 
      \left[ \frac{1+z}{10} \frac{0.15}{\Omega_m h^2} \right] \left[ \frac{\Omega_b h^2}{0.023} \right]\,{\rm mK} ~.
\end{align}
Here, $T_\gamma$ and $T_S$ denote the temperature of the cosmic microwave background (CMB) radiation and 
the spin temperature of the intergalactic neutral hydrogen, respectively. The term $d v_\parallel/d r_\parallel$ represents the peculiar velocity gradient of the gas along the LOS, and it is responsible for 
introducing the RSD effect as will be detailed in the next section.

The spin temperature term in Equation~(\ref{eq:tbeq}) traces the CMB temperature and the thermal history of the IGM via different coupling mechanisms. These include the collisional (thermal) coupling of the intergalactic gas, as well as radiative coupling effects through CMB and Ly$\alpha$ photons. The coupling through resonant scattering of Ly$\alpha$ photons is known as the Wouthuysen-Field (WF) effect, and it is sourced by the cosmic soft-UV Ly$\alpha$ background described in the previous section. Taking into account these processes, we can express the offset $1-T_\gamma/T_S$ as 
\begin{equation}
    1-\frac{T_\gamma}{T_S} = \frac{x_c + x_\alpha}{1 + x_c + x_\alpha} \left( 1-\frac{T_{\gamma}}{T_K} \right)~,
\end{equation}
where $T_K$ is the kinetic temperature of the intergalactic gas, and $x_c$ and $x_\alpha$ are the collisional and WF radiative coupling coefficients, respectively. While the Ly$\alpha$ background is the responsible for the WF coupling as mentioned above, the heating of the intergalactic gas is mostly dominated by the X-ray background. 

For the calculation of the 21 cm signal, LIMFAST follows the methodology developed 
in 21cmFAST, except for the Ly$\alpha$ background derivation. As mentioned in the previous section, for the Ly$\alpha$ background we adopt our metallicity-dependent SEDs, which results in a varying Ly$\alpha$ background due to the evolution of metallicity, in addition to that of star formation, with redshift. For the X-ray background, we use the prescriptions relating luminosity and star formation adopted by \cite{Park2019}. Finally, for both radiation background calculations, we have modified the original code to take into account our star formation formalism.

\subsection{Redshift-Space Distortions} \label{sec:rsd}

The implementation of RSD calculations in LIMFAST generally follows that of 21cmFAST, as described in \cite{Mesinger2011} and \cite{GM2018}, which includes the numerical approach to transform the data from real to redshift space outlined by \cite{Jensen2013}. This method divides each simulation cell into sub-cells of equal intensity and computes their new position according to the peculiar velocities. Then, the sub-cells are re-grid to the original resolution \citep[see Section 3.1.2 in][]{Jensen2013}. Finally, for the calculation of intensities with RSD, we use the default 21cmFAST cut-off velocity gradient value of ${dv_\parallel}/{dr_\parallel}=0.2\, H(z)$. This limit is set to avoid extremely large intensities when the velocity gradient presents large values. The exact value for this cut-off is somewhat arbitrary, but \cite{Mao2012} found that values around the default one used here do not produce extreme departures from the true results (their Section 9), although this depends on the specific simulation and case. We have verified that using ${dv_\parallel}/{dr_\parallel}=0.1 \,H(z)$ or ${dv_\parallel}/{dr_\parallel}=0.3\, H(z)$ does not significantly change our results. This cut-off, however, is not required and, therefore, not used in the 21 cm calculations. For the 21 cm line, the full radiative transfer derivation ignoring the optically thin assumption results in terms that vanish instead of going to infinity \citep[see Section 5.1.1 in][]{Mao2012}. In principle, a similar full derivation would be possible for other lines, but this requires accounting for the specific radiative processes affecting each line of interest, which goes beyond the scope of LIMFAST in its current form. We leave these calculations to future work and here simply use the cut-off method for the star-formation lines. For interested readers, a more detailed summary of the RSD-related calculations in LIMFAST and effects on the LIM signals of interest is provided in Appendix~\ref{sec:psrsd}.

\section{Results}\label{sec:results} 

We present in the remainder of this section the results from LIMFAST runs considering simulation boxes of 0.5\,Gpc on a side, with a cell size of $2.5^3\,\mathrm{Mpc}^3$, from $z=15$ to $z=5$. In this paper, we focus on the 21 cm line tracing the neutral IGM and various optical/UV lines generally emitted from \ion{H}{2} regions created by the formation of massive stars in galaxies. These star-formation lines are some of the strongest emission features in the optical/UV spectrum of a typical star-forming galaxy, including Ly$\alpha$, H$\alpha$, H$\beta$, [\ion{O}{2}] $3727 \angs$, and [\ion{O}{3}] $5007 \angs$. It is noteworthy that the intensity field of Ly$\alpha$ (and in principle other hydrogen recombination lines like H$\alpha$, though to a much less extent) is not only sourced by recombinations of ionized gas inside galaxies; it also contains emission from the diffused, ionized IGM and the redshifted background radiation associated with sources within the ``horizons'' \cite[see e.g.,][]{Furlanetto_2006PhR, HF_2012}. These additional components are self-consistently calculated and generated separately. In \citetalias{Sun_2022P2}, we present extensions that model emission lines from the neutral ISM, such as the [\ion{C}{2}] $158$\,$\mu$m and CO rotational lines. For predictions displayed in this section, we present the fiducial case where we adopt a fixed value of the ionization parameter $\log U = -2$ consistent with the enhanced $U$ values typically observed for high-$z$ galaxies \citep{Sanders2016,Strom2017,Katz_2023}. A plausible variation that assumes metallicity-dependent $U$ is also considered. We will revisit these assumptions later again in Section~\ref{sec:discuss:uvsz} and discuss their implications for future LIM observations. In Section \ref{sec:evo}, we present the results for the redshift evolution of the variety of quantities of interest, such as the SFRD, metallicity, IGM neutral fraction, and mean line intensities. In Section \ref{sec:ps}, we show power spectra derived from the simulated line intensity fields. Results from the literature for are also displayed for comparison, and we discuss the origin of several key differences between our results and previous studies in Section~\ref{sec:discussion}.

\subsection{Redshift Evolution of the Cosmic Means}\label{sec:evo} 

\subsubsection{Star Formation Rate Density}\label{sec:sfrd} 

\begin{figure}
 \centering
 \includegraphics[width=0.48\textwidth]{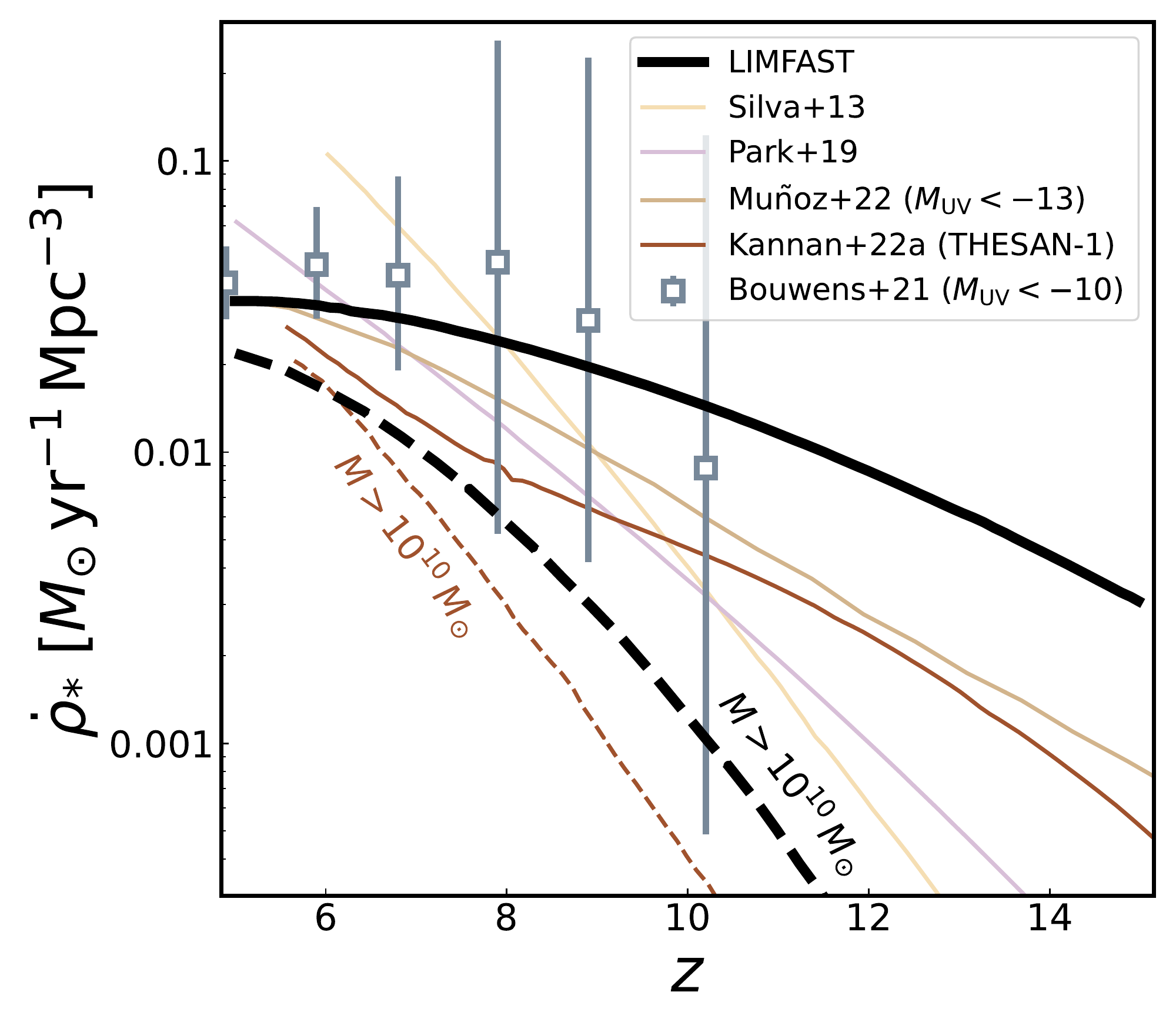}
 \caption{The evolution of the cosmic star formation rate density computed by LIMFAST with the fiducial model described in previous sections. For comparison, we also plot similar results from the literature \citep{Silva2013, Park2019, Munoz2022, Kannan_2022_THESAN} as colored lines, together with observational data from \cite{Bouwens2021} after extrapolating the luminosity functions to a limiting magnitude of $M_{\rm UV} = - 10$. The error bars represent the 16th and 84th percentiles derived from the Schechter parameters reported. The dashed lines show the fractions of cosmic star formation associated with massive halos with $M>10^{10}\,M_{\odot}$.}
 \label{fig:sfrdz}
\end{figure}

By averaging the SFRDs of individual cells computed as explained in Section~\ref{sec:sfxs} over the coeval box at different redshifts, we can determine the mean cosmic SFRD evolution. In Figure \ref{fig:sfrdz}, we show results from LIMFAST, assuming the fiducial model and default setups specified in Section~\ref{sec:code} (black solid line). The colored lines illustrate a compilation of predictions from the literature \citep{Silva2013,Park2019,Munoz2022,Kannan_2022_THESAN}, which typically involve different assumptions of $M_\mathrm{min}$ that can host star-forming galaxies. Unless otherwise labeled, the atomic cooling limit is assumed, as is the case for LIMFAST (see Section~\ref{sec:sfxs}). These theoretical predictions are compared with the data points that represent SFRDs implied by the observed UVLFs \citep{Bouwens2021}. Note that to make fair comparisons to our results, we use the Schechter parameters (including their variances but not covariances) reported in \citet{Bouwens2021} to extrapolate and integrate the UVLF measurements down to a magnitude of $M_{\rm UV} = - 10$, which roughly corresponds to a halo mass at the atomic cooling limit in the redshift range of interest. 

Overall, the SFRD evolution from LIMFAST agrees well with the latest observations after plausible extrapolations, although these constraints become rather weak beyond $z \sim 7$. Our SFRD is generally comparable (within a factor of a few) to models from the literature at $5 \lesssim z \lesssim 8$, with the difference in the shape and amplitude growing larger towards higher redshift. As will be further discussed in Section~\ref{sec:vs21cmfast}, the overall higher amplitude and shallower slope are mainly because our model allows star formation to occur in low-mass halos (i.e., faint galaxies) down to the atomic cooling limit with a non-negligible efficiency. To demonstrate this, we also include in Figure \ref{fig:sfrdz} cases showing the SFRD from only the massive halos with $M>10^{10}\,M_{\odot}$.

\subsubsection{Metallicity} 

\begin{figure}
 \centering
 \includegraphics[width=0.48\textwidth]{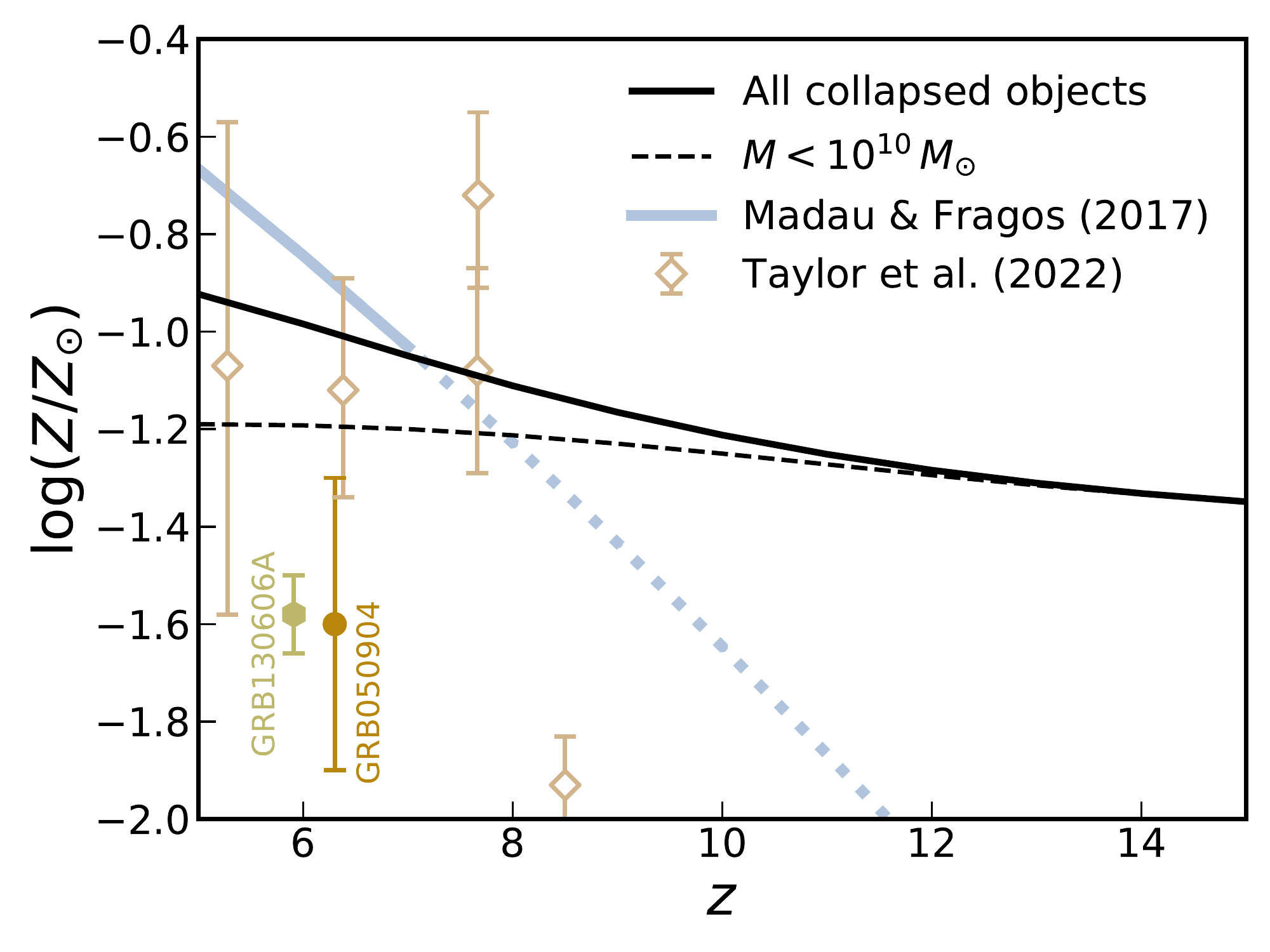}
 \caption{Metallicity evolution of the collapsed objects in LIMFAST, as derived from Equation (\ref{eq:metal}) considering all halos more massive than $M_\mathrm{min}$ (black solid curve) or those with $M_\mathrm{min}<M<10^{10}\,M_{\odot}$ (black dashed curve). The data points represent the metallicity estimates of five $z>5$ galaxies measured from JWST/NIRSpec data using the direct method \citep{Taylor2022}, as well as those of the two highest redshift GRB hosts to date with confident, absorption-based metal detections \citep{Thone2013,Bolmer2019}. The blue curve shows a best-fit metallicity evolution of the observed galaxy population as a whole (extrapolated to $z>7$) adopted from \citet{MadauFragos2017}.}
\label{fig:metallicity}
\end{figure}

Similar to the SFRD, we can compute the mean cosmic metal enrichment history by averaging the metallicity per cell over coeval boxes. The solid black line in Figure \ref{fig:metallicity} shows the mean metallicity evolution of the collapsed structures, following Equation~(\ref{eq:metal}). Specifically, these metallicity values 
represent the mean metallicities of the gas (and stars) residing in galaxies in the simulation cells at different redshifts, but they should not be confused with the mean halo metallicity. For comparison, a number of metallicity estimates of $z>5$ star-forming galaxies and gamma-ray burst (GRB) hosts available to date, together with the metal enrichment history and its extrapolation to $z>7$ implied by the cosmic star formation history \citep{MadauFragos2017}, are also shown. The galaxy metallicities are determined from recent analysis of JWST/NIRSpec \citep{Taylor2022} using the direct $T_{e}$ method, whereas the GRB metallicities are measured from absorption spectra of their afterglows \citep{Thone2013,Bolmer2019}. Because GRBs may preferentially reside and thus represent the ISM of low-mass (i.e., metal-poor) galaxies \citep[e.g.,][and references therein]{Graham2017,Palmeiro2019}, we also show the metallicity evolution considering only halos with $M<10^{10}\,M_{\odot}$. Given the small sample size and large uncertainties associated with the metallicity estimates, especially for GRBs where the analysis is highly sensitive to properties of their host halos \citep[see, e.g.,][]{Metha2021}, we find broad agreement between these observational constraints/implications and our model predictions. 

\subsubsection{Neutral Fraction}\label{sec:xhi} 

\begin{figure}
 \centering
 \includegraphics[width=0.48\textwidth]{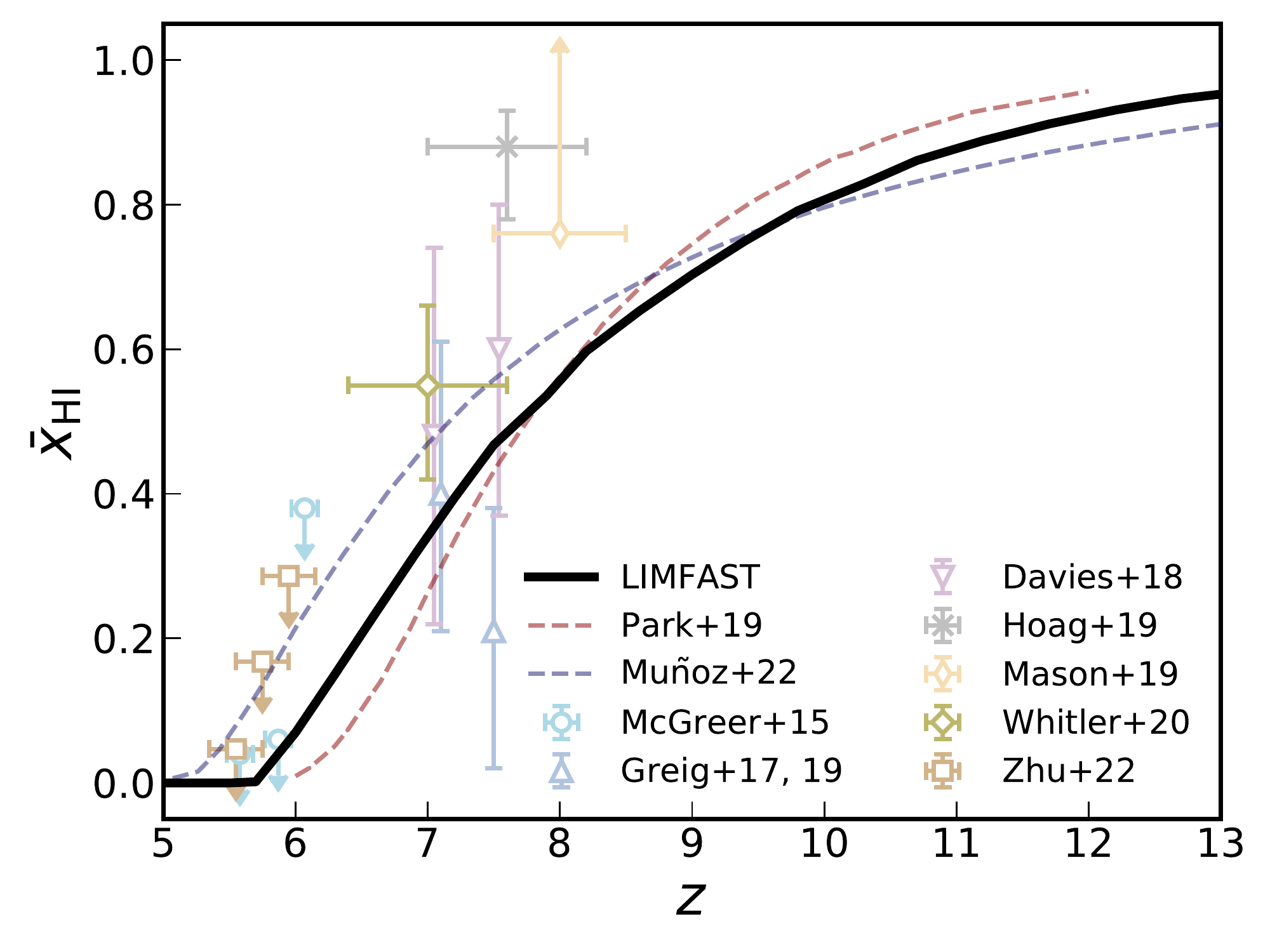}
 \caption{Evolution of the volume-averaged IGM neutral fraction $\bar{x}_\mathrm{HI}$ as computed by LIMFAST with a 10\% escape fraction of ionizing photons (black solid curve). For comparison, the colored data points show constraints from the literature based on different methods, including quasar absorption spectra \cite[][]{McGreer2015,Greig2017,Greig2019,Davies2018,Zhu2022} and Ly$\alpha$ emission \citep{Hoag2019, Mason2019, Whitler2020}. Furthermore, the dashed curves represent $\bar{x}_\mathrm{HI}$ evolution predicted by other previous 21cmFAST-based simulations \citep{Park2019,Munoz2022} that also broadly match observational constraints available to date.}
 \label{fig:xhi}
\end{figure}

The redshift evolution of the IGM neutral fraction is a key diagnostic for our simulations of the reionization process. Figure \ref{fig:xhi} shows the evolution of the gas neutral fraction as computed by LIMFAST, assuming a LyC escape fraction of 10\% that yields a CMB optical depth $\tau_\mathrm{CMB} \approx 0.06$ consistent with recent estimates (see also \citetalias{Sun_2022P2} for a comparison of the reionization scenarios and corresponding optical depths from plausible variations of the stellar feedback). The colored data points represent observational constraints from quasar absorption spectra \citep{McGreer2015,Greig2017,Davies2018,Greig2019,Zhu2022} and Ly$\alpha$ transmission \citep{Hoag2019,Mason2019,Whitler2020}, whereas the colored lines denote some alternative models from previous 21cmFAST-based simulations \citep{Park2019, Munoz2022}. We note that some recent observations have provided strong evidence for a late completion of reionization by $z\lesssim5.5$ \cite[e.g.,][]{Zhu2022}, which poses challenges on existing reionization models. As our fiducial model still agrees with the updated upper limits at $z<6$, we do not attempt to match them and consider scenarios producing $\tau_\mathrm{CMB}$ values consistent with CMB observations to be generally valid.

\begin{figure*}
 \centering
 \includegraphics[width=0.95\textwidth]{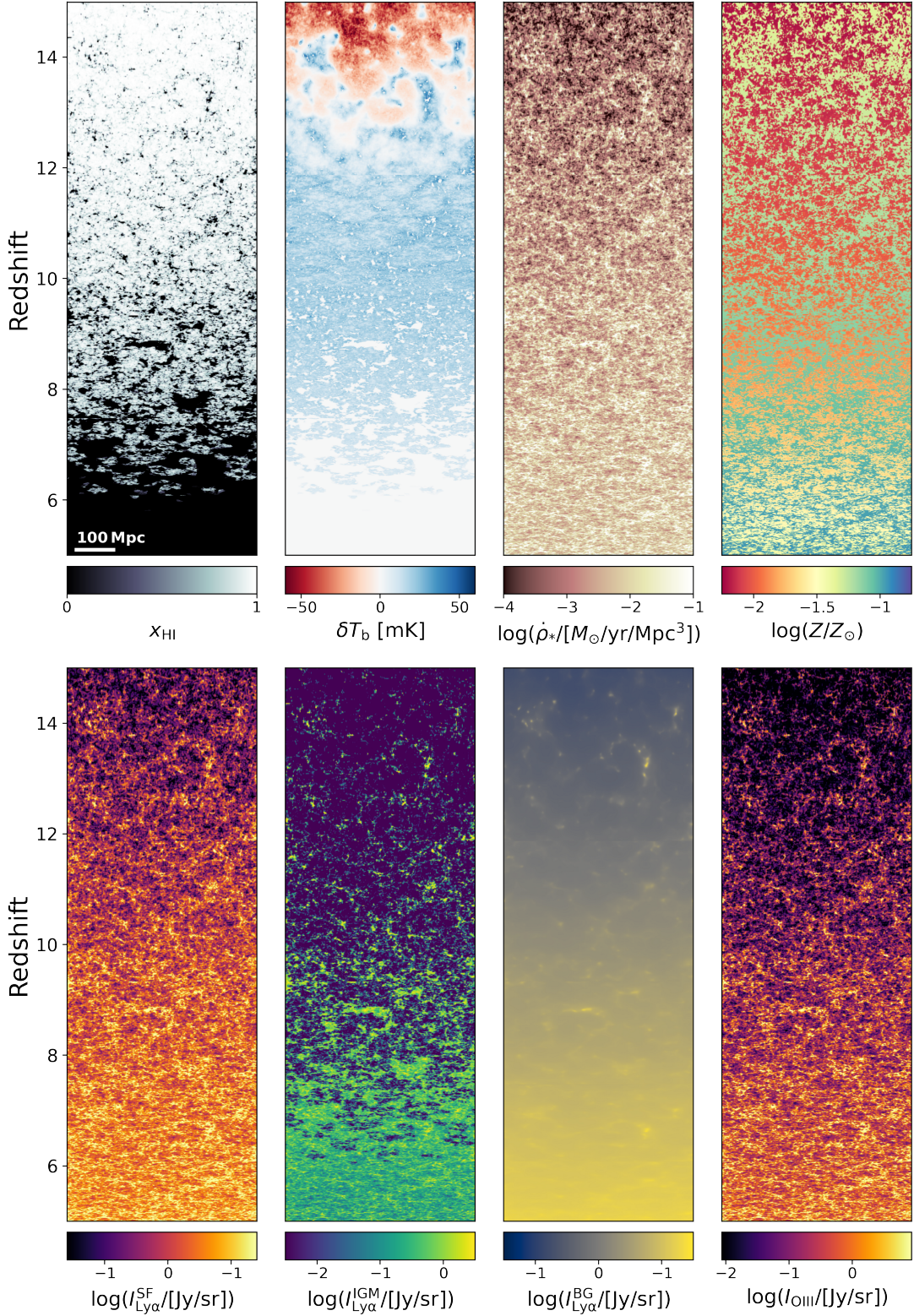} 
 \caption{LIMFAST light cones over redshift $5 \le z \le 15$. The 8 panels shown correspond to the hydrogen neutral fraction, the brightness temperature of the 21 cm line, the SFRD, the metallicity of collapsed structures, and the intensities of Ly$\alpha$ (from star formation, the ionized IGM, and the scattering background, respectively) and [\ion{O}{3}] $5007 \angs$ emission. The color schemes are identically normalized (to the maximum value) for [\ion{O}{3}] and Ly$\alpha$ from star formation to highlight their difference.}
 \label{fig:lightcones}
\end{figure*}

\subsubsection{Line Emission}\label{sec:lumevo} 

To showcase the output of LIMFAST, we show in Figure \ref{fig:lightcones} the LIMFAST light cones for various quantities. As in 21cmFAST, these light cones are created by first generating coeval boxes at discrete redshift steps over the redshift range of interest and then interpolating (after being rotated to avoid repeating structures) between redshifts along the LOS to describe the continuous evolution of each quantity. From the top to the bottom row, the signals displayed are the neutral fraction of hydrogen, the 21 cm brightness temperature, the SFRD, the metallicity of collapsed structures, Ly$\alpha$ intensities from star formation, recombinations in the diffuse ionized IGM, and the scattering of UV photons in the IGM, as well as the [\ion{O}{3}] $5007 \angs$ line intensity, respectively. 

Figure \ref{fig:lightcones} visualizes qualitatively the correlation between different quantities. For instance, beyond the well-studied correlation of the 21 cm signal with the hydrogen neutral fraction, it is easy to see the increase of metallicity (metal enrichment) that corresponds to the SFRD evolution, which is directly traced by the evolving Ly$\alpha$ emission from star-forming galaxies. The Ly$\alpha$ emission from the diffused ionized IGM, on the other hand, shows a stronger dependence on the ionization state of the IGM than the amount of cosmic star formation, and its small-scale spatial fluctuations appear to be weaker compared to the star-formation component of Ly$\alpha$ emission. Small-scale fluctuations are even weaker for the Ly$\alpha$ background radiation from the scattering of UV photons, due to the large Ly$\alpha$ ``horizon distance'' (a few hundred comoving Mpc at $z\sim10$) associated with the distance beyond which UV photons can no longer redshift into the Ly$n$ series and produce Ly$\alpha$ radiation \citep{HF_2012}. Finally, compared with hydrogen lines, metal lines like [\ion{O}{3}] $5007 \angs$ are much more sensitive to the metallicity, which controls the hardness of the stellar ionizing spectrum. As a result, the non-trivial metallicity dependence leads to a stronger redshift evolution for the [\ion{O}{3}] line compared with Ly$\alpha$, as shown by the steeper color gradient. 

\begin{figure*}
 \centering
 \includegraphics[width=0.48\textwidth]{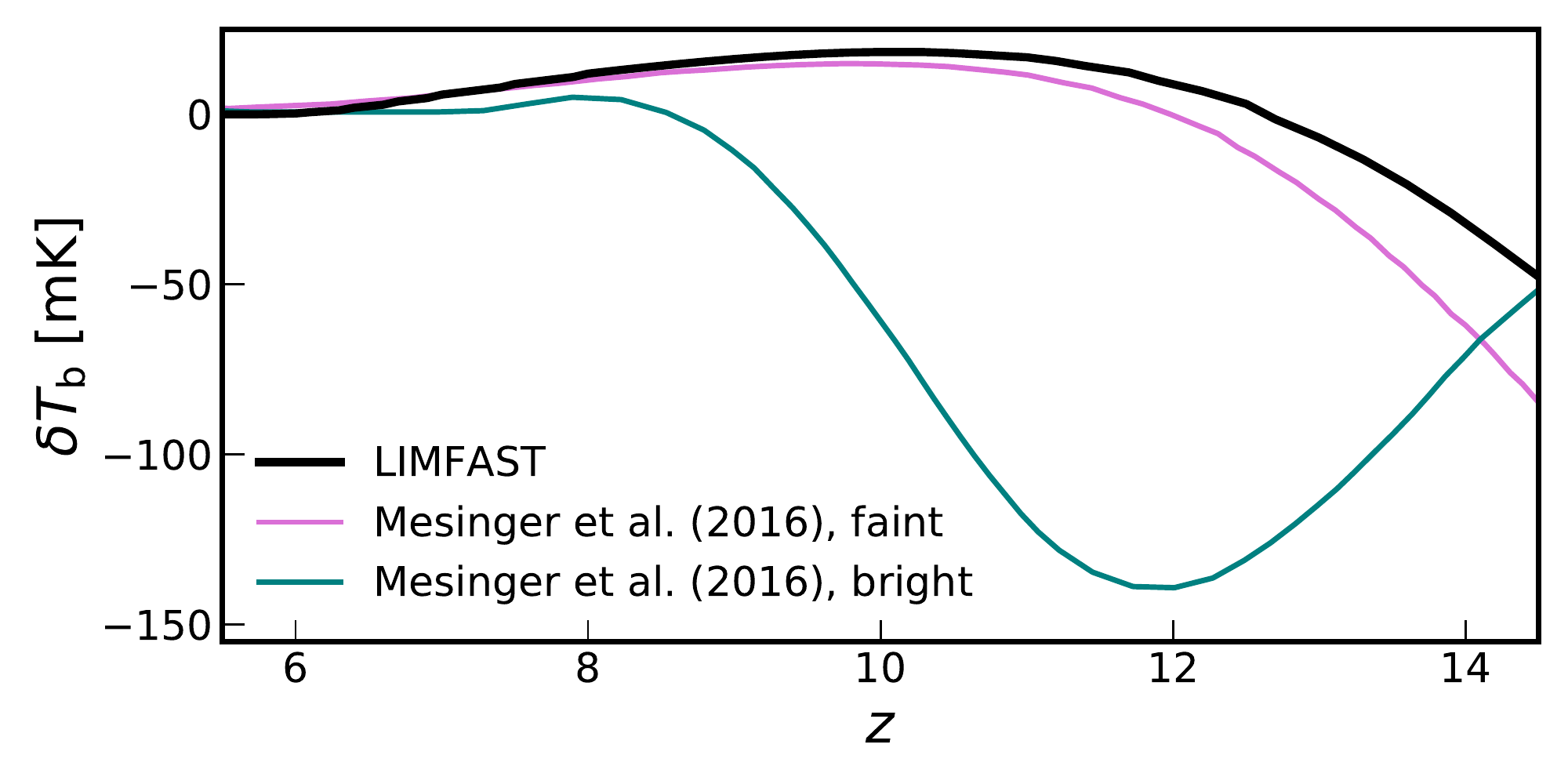} \includegraphics[width=0.48\textwidth]{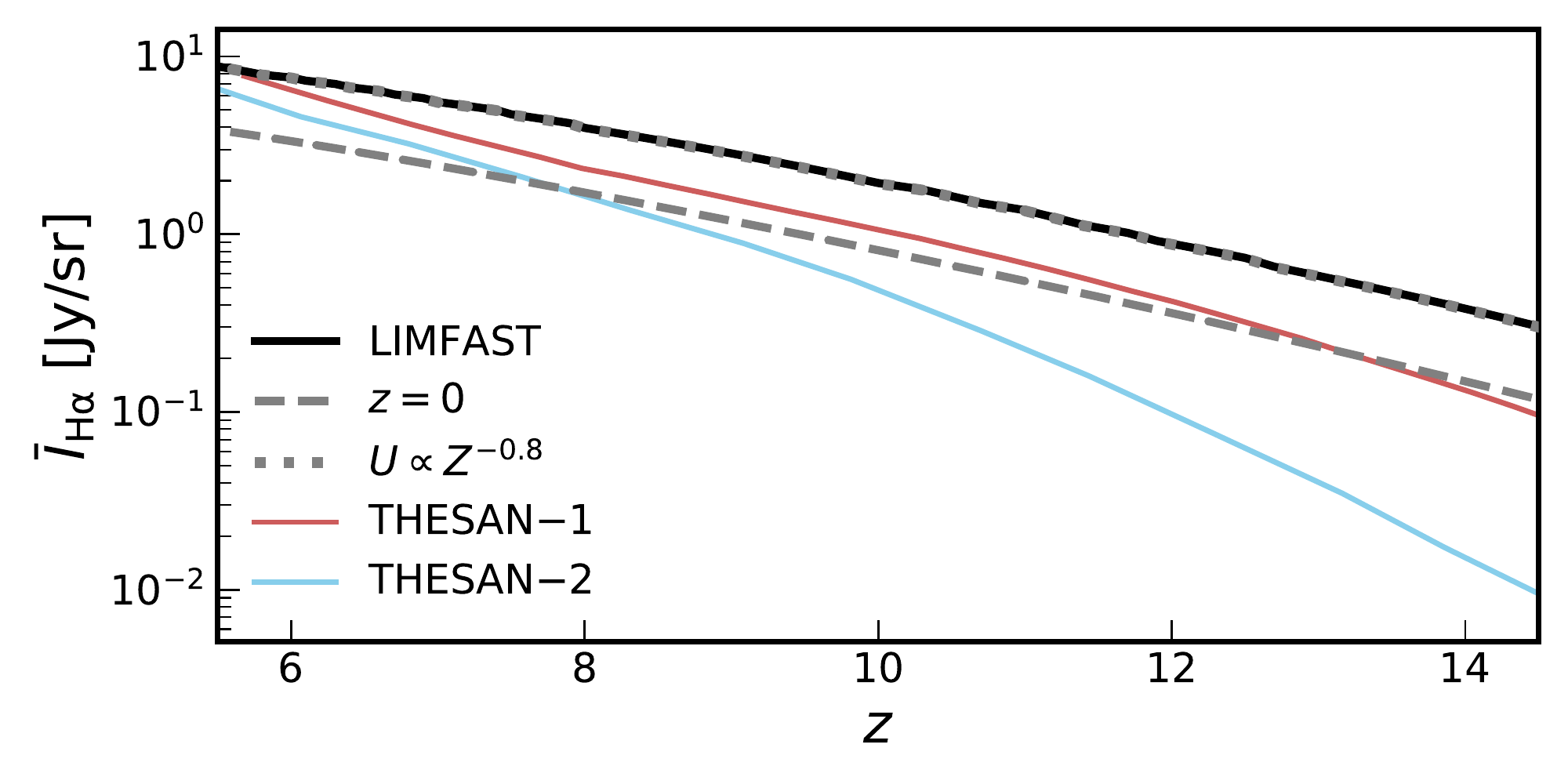}
 \includegraphics[width=0.48\textwidth]{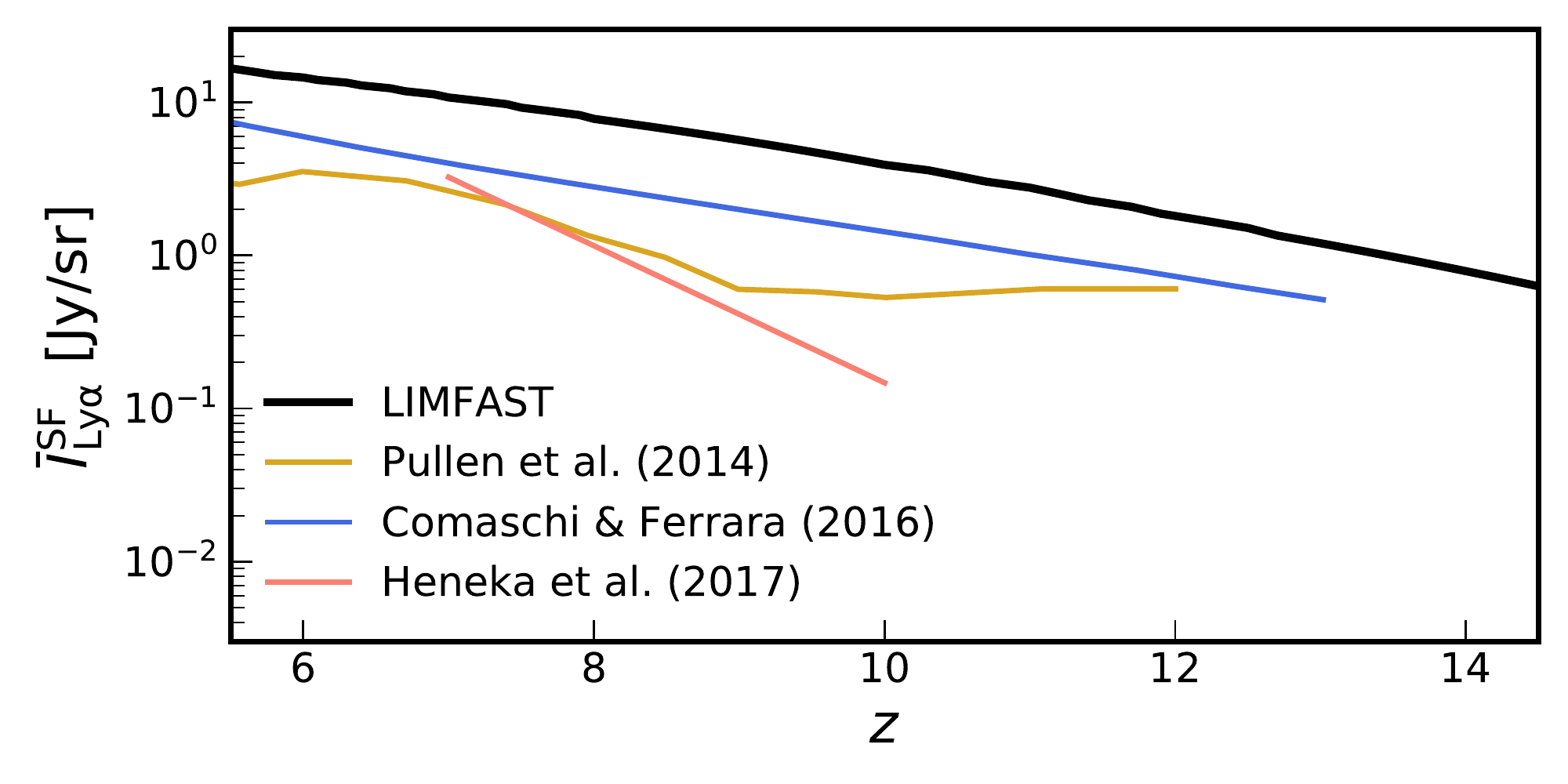} \includegraphics[width=0.48\textwidth]{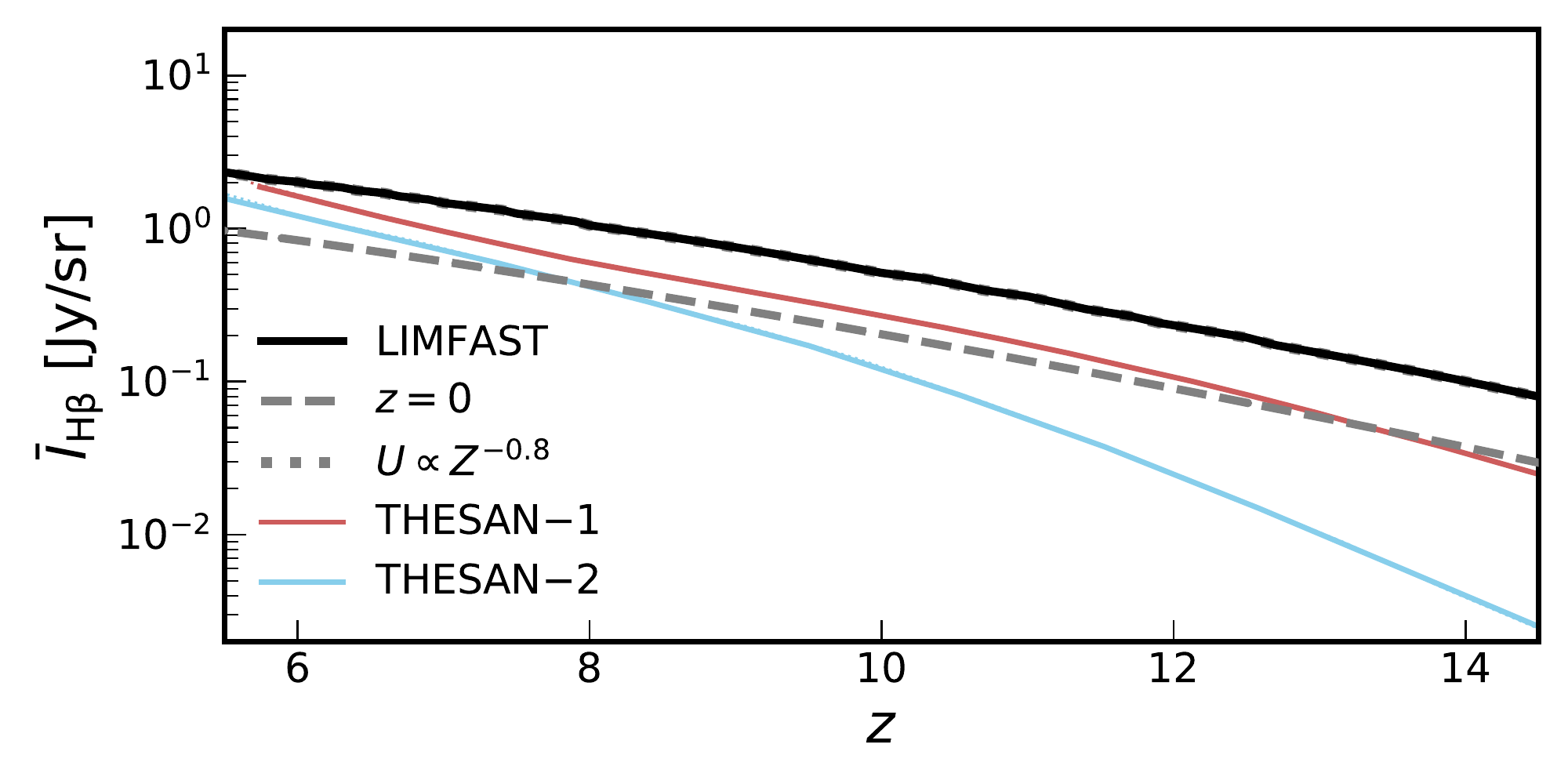}
 \includegraphics[width=0.48\textwidth]{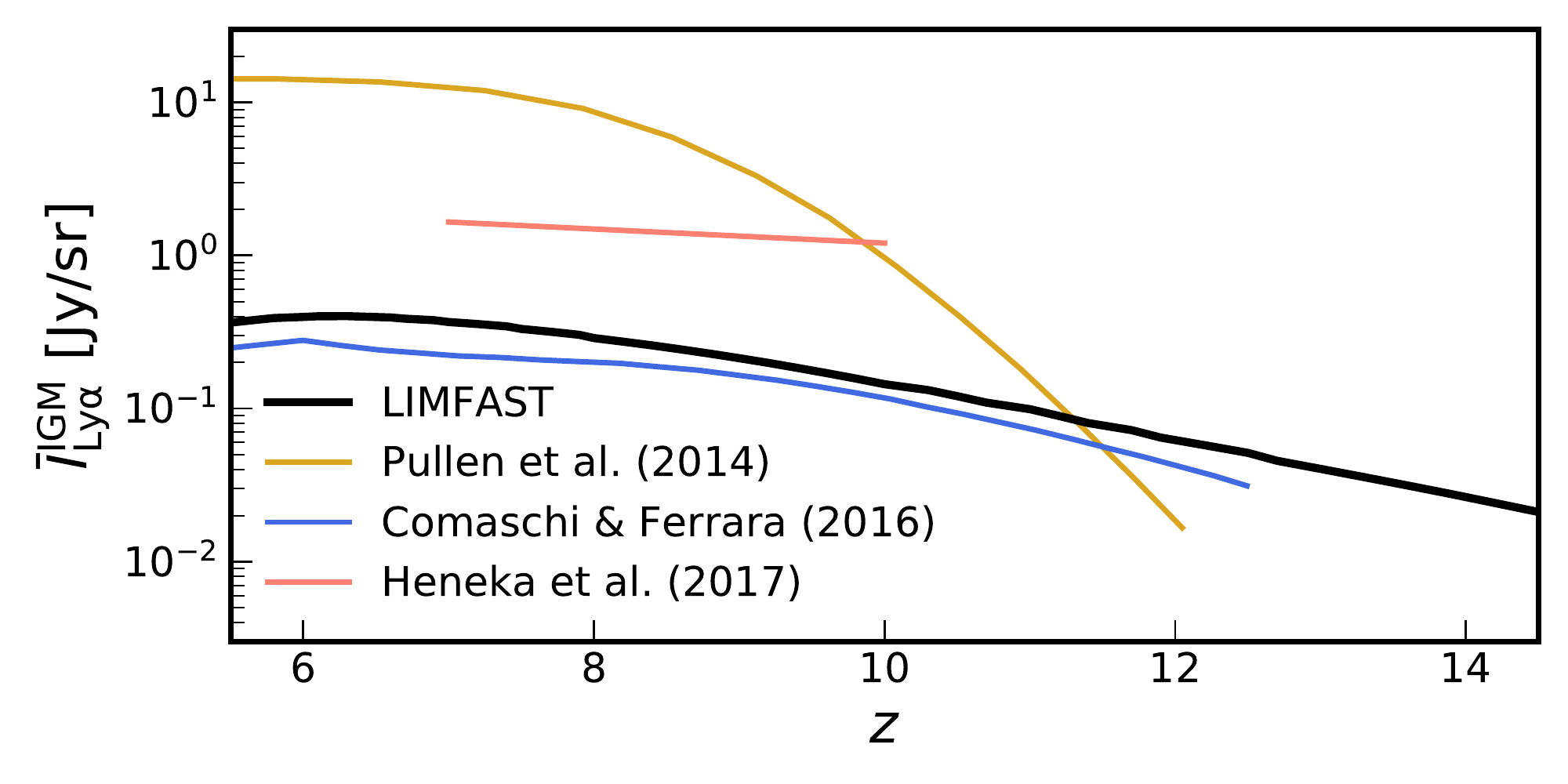} \includegraphics[width=0.48\textwidth]{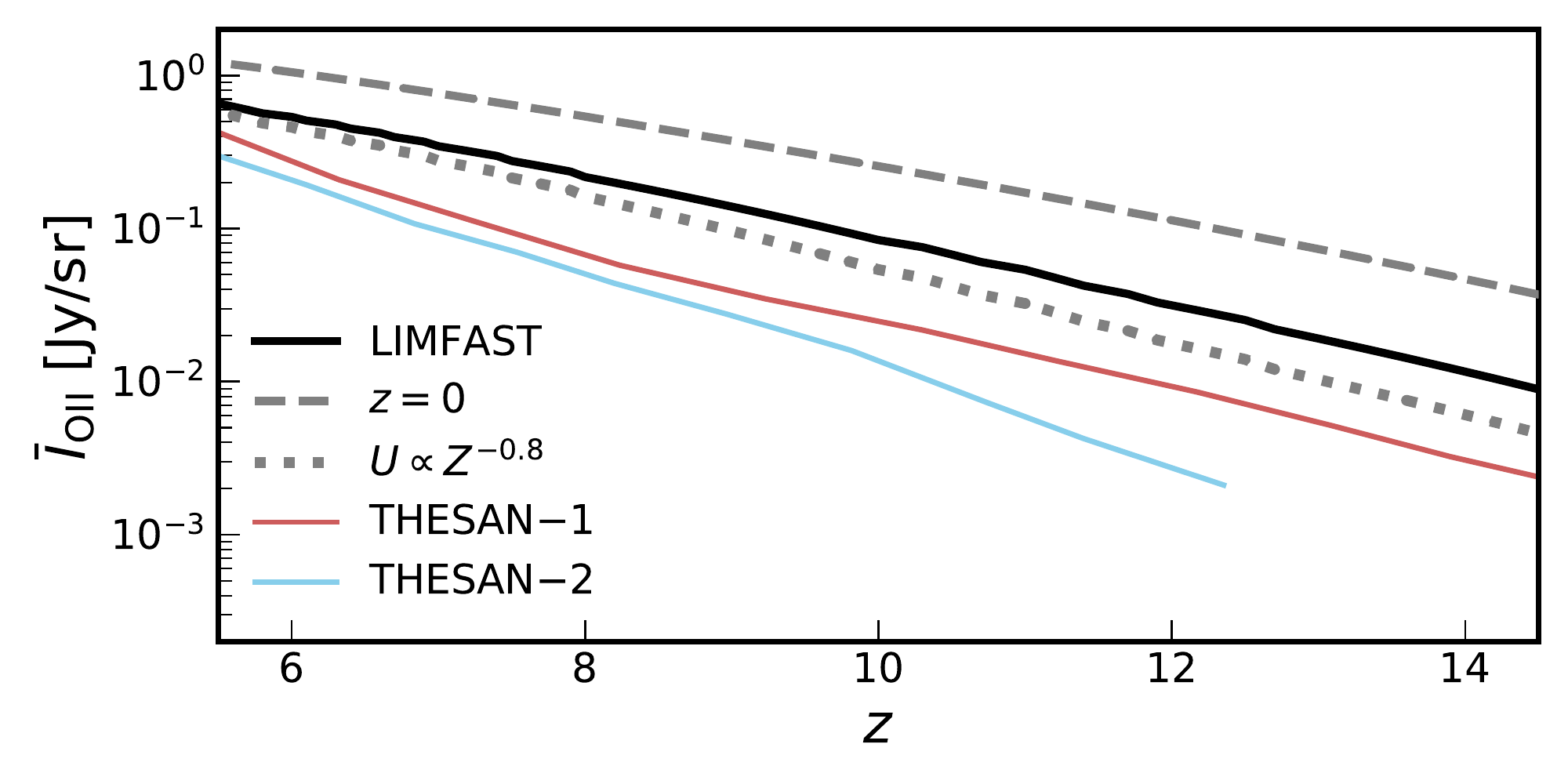}
 \includegraphics[width=0.48\textwidth]{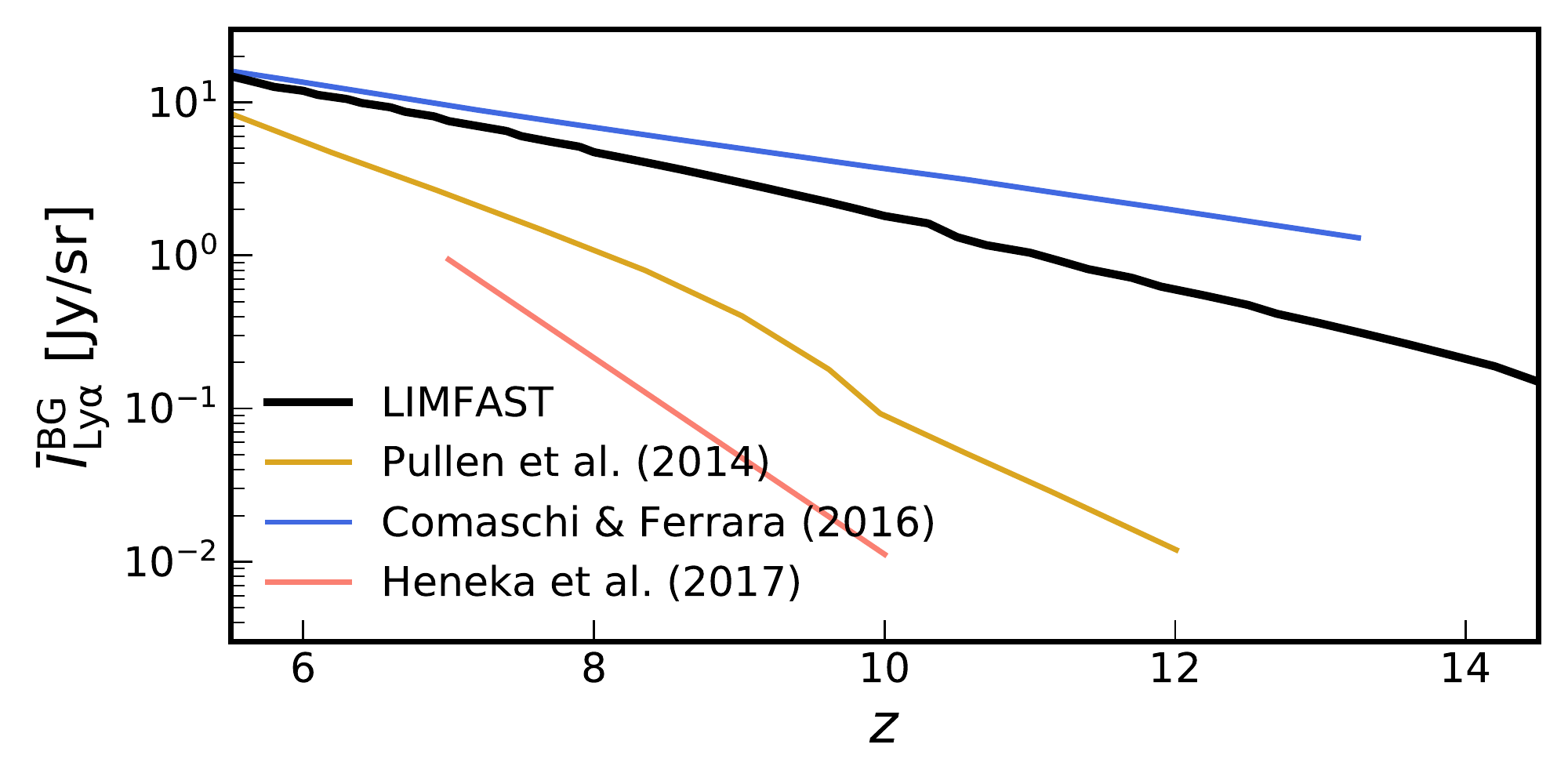} \includegraphics[width=0.48\textwidth]{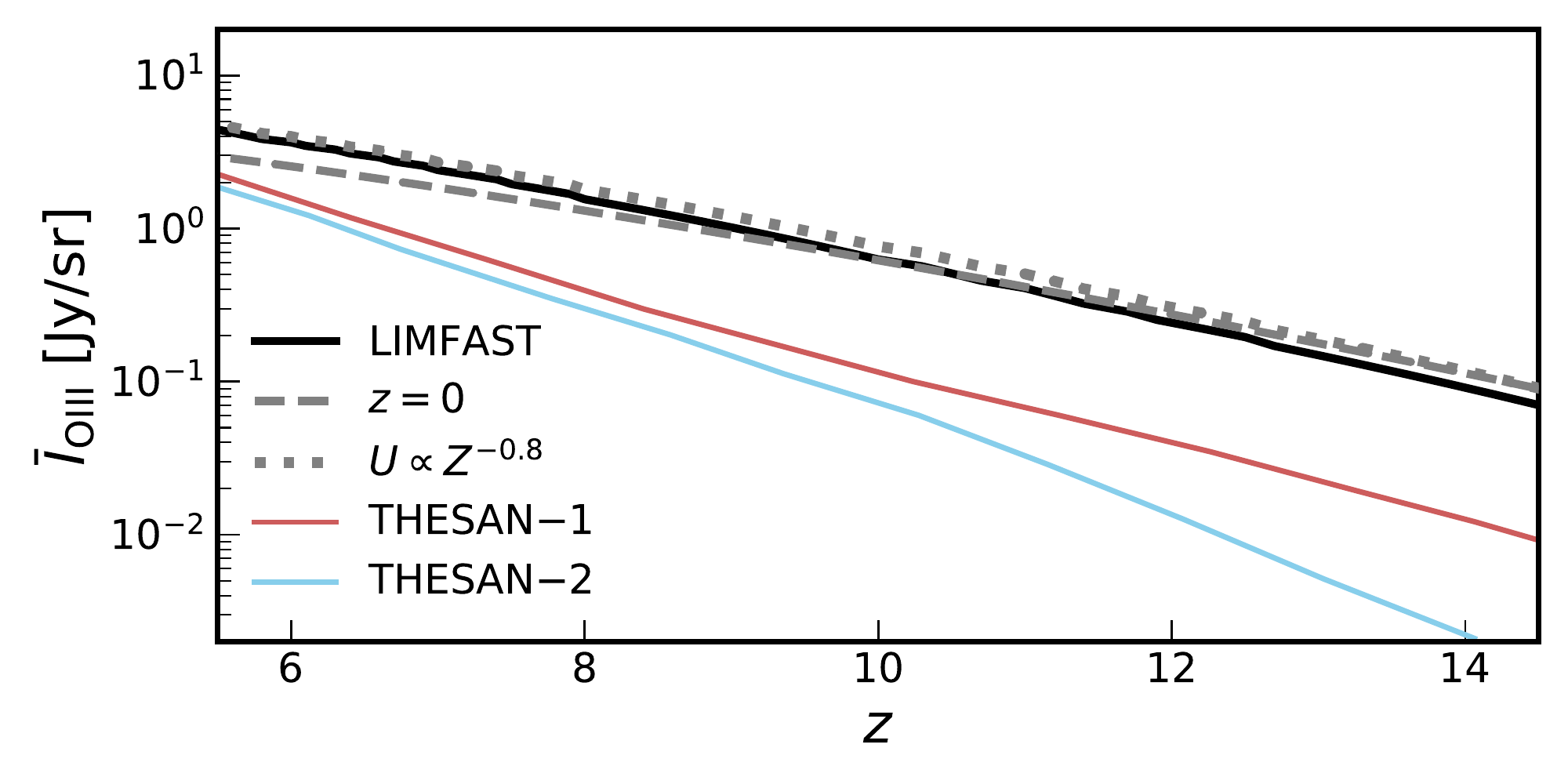}
 \caption{Redshift evolution of the brightness of emission lines computed by LIMFAST (solid black curves), compared with results from the literature shown by the colored curves \citep{Pullen2014,Comaschi2016,Mesinger2016,Heneka2017,Kannan_2022_LIM}. Two alternative cases assuming local scaling relations and the anti-correlation $U \propto Z^{-0.8}$ for the Balmer and oxygen lines are shown by the dashed and dotted curves in gray, respectively.}
 \label{fig:lines}
\end{figure*}

A more quantitative view of the redshift evolution of these LIM signals is provided in Figure \ref{fig:lines}, where we also show results from the literature for comparison. In addition to our fiducial model, for the Balmer and oxygen lines we consider two alternative cases that exemplify how variations of the photoionization modeling affect the predicted mean intensity evolution of different lines. As denoted by the gray dashed lines, one of such alternative cases is to adopt a local, redshift-independent relation between the line luminosity and the SFR, which otherwise evolves with redshift through the metallicity dependence in our fiducial model. Here, we take scaling relations similar to those in \citet{Gong2017}, which are calibrated against galaxies observed in the local universe. Specifically, considering the scaling relation $L_\mathrm{line} = C\, \dot M_*$, we have $C = 1.3 \times 10^{41} \, \mathrm{erg\,s^{-1}}\,\dot M_\star^{-1}$ for H$\alpha$ \citep{Kennicutt1998}, $C = 4.4 \times 10^{40} \, \mathrm{erg\,s^{-1}}\,\dot M_\star^{-1}$ for H$\beta$ from the relation ${\rm H}\alpha/{\rm H}\beta=2.86$, $C = 7.2 \times 10^{40} \, \mathrm{erg\,s^{-1}}\,\dot M_\star^{-1}$ for [\ion{O}{2}] from the relation [\ion{O}{2}]$/{\rm H}\alpha=0.57$ \citep{Kennicutt1998}, and $C = 1.3 \times 10^{41} \, \mathrm{erg\,s^{-1}}\,\dot M_\star^{-1}$ for [\ion{O}{3}] \citep{Ly2007}. Besides the clear impact on the overall normalization of the redshift evolution, assuming locally-calibrated scaling relations also results in a slightly shallower slope, especially for the metal lines, because in this case the ionizing spectrum does not harden as the metallicity decreases (towards higher redshift).

The other alternative case, as denoted by the gray dotted lines, assumes that the ionization parameter $U$ anti-correlates with metallicity following $U \propto Z^{-0.8}$, instead of assigning a fixed value like $\log U=-2$ as in the fiducial case. Such an anti-correlation is broadly in line with the ISM physical condition of high-$z$ galaxies probed by nebular emission line ratios \citep{Sanders2016} and the expectation of simple analytic models of wind-driven bubble \citep{Dopita2006}. Due to the way the oxygen line ratio $\mathrm{O32 = [O\,III]/[O\,II]}$ depends on $U$, the assumption of $U \propto Z^{-0.8}$ leads to a steeper (shallower) redshift evolution of $\bar{I}_\mathrm{O\,II}$ ($\bar{I}_\mathrm{O\,III}$), though the difference from the fiducial case with $\log U=-2$ is modest. While neither a universal, fixed $U$ nor the exact anti-correlation considered may be true in reality, they illustrate how much LIM predictions might be affected by the detailed photoionization modeling of \ion{H}{2} regions associated with the formation of massive stars. Without a good understanding of the chemical abundance and ionization conditions of \ion{H}{2} regions in high-$z$ galaxies, it would therefore be difficult to gauge how reliably LIM measurements of lines like [\ion{O}{2}] and [\ion{O}{3}] can be applied to constraining the cosmic SFRD at high redshift \citep{Gong2017,Sun2022}. We will further elaborate on this and its implications for future LIM observations in Section~\ref{sec:discuss:uvsz}. 

Qualitatively, the redshift evolution of the various lines extracted from our simulations share a few similarities with results in the literature. For line intensities that directly trace the ongoing cosmic SFRD (including the accumulated Ly$\alpha$ background that encodes the past star formation history), the steepness of the redshift evolution is comparable to other works (e.g., \citealt{Comaschi2016} and THESAN-1) that allow stars to form efficiently in low-mass halos, in such a way that consistently predicts the faint-end slope of the UVLFs observed. The calibration against the UVLF, typically done at at $z\sim5$--6, also leads to reasonable agreement in the amplitude at these redshifts. Quantitatively, discrepancies ranging from orders of a few at $z\sim5$ to several hundred at $z>10$ exist, as a result of different assumptions made for galaxy formation and additional effects like dust attenuation. Impacts of systematic differences caused by the higher and slowly evolving SFRD and the exclusion of attenuation corrections, as well as complications due to the photoionization modeling that lead to deviations from the local relations \citep[e.g.,][]{Silva2017,Kewley2019}, will be further discussed in Section~\ref{sec:discussion:lit}.

\subsection{Power Spectra}\label{sec:ps} 

\begin{figure*}
 \centering
 \includegraphics[width=0.24\textwidth]{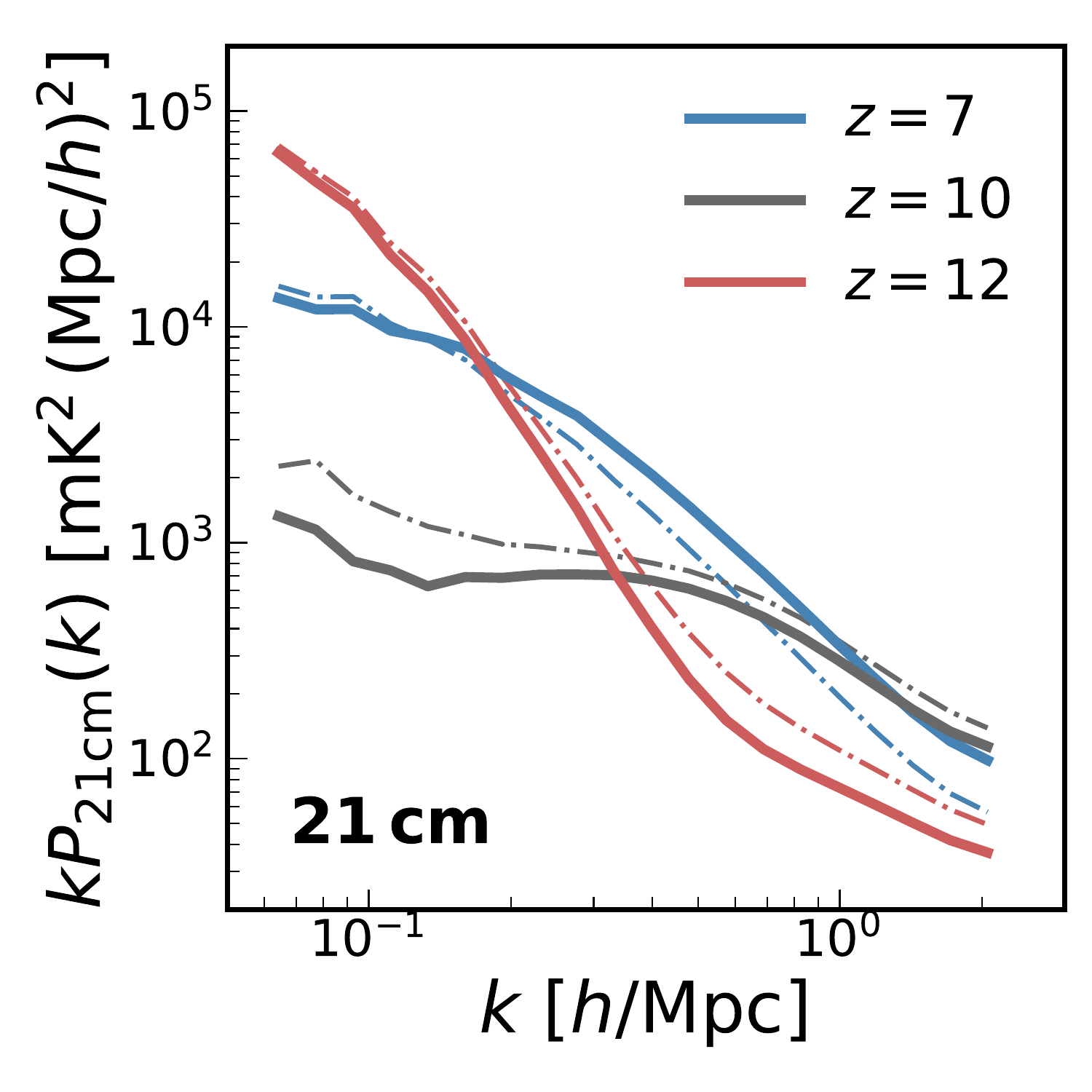} \includegraphics[width=0.24\textwidth]{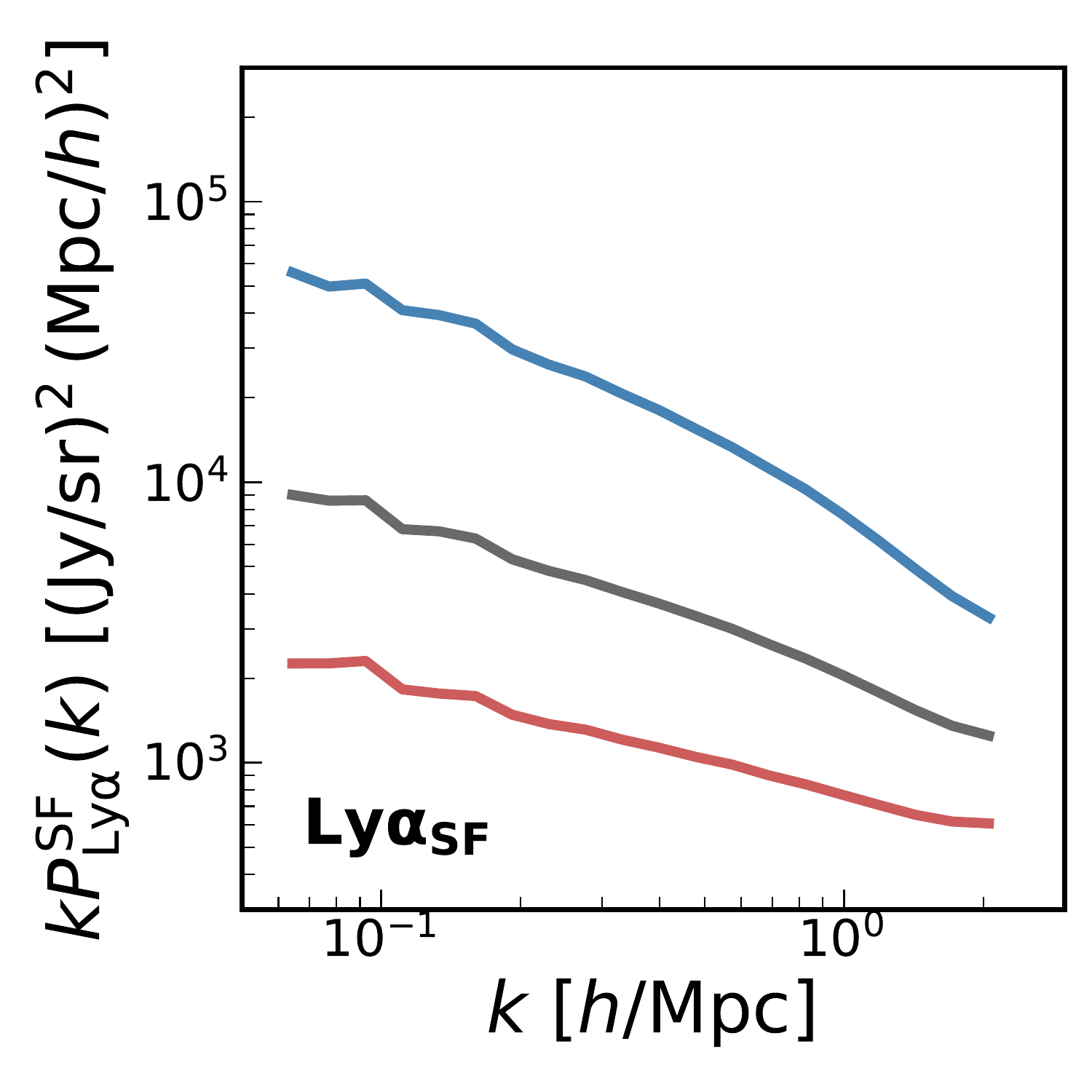}
 \includegraphics[width=0.24\textwidth]{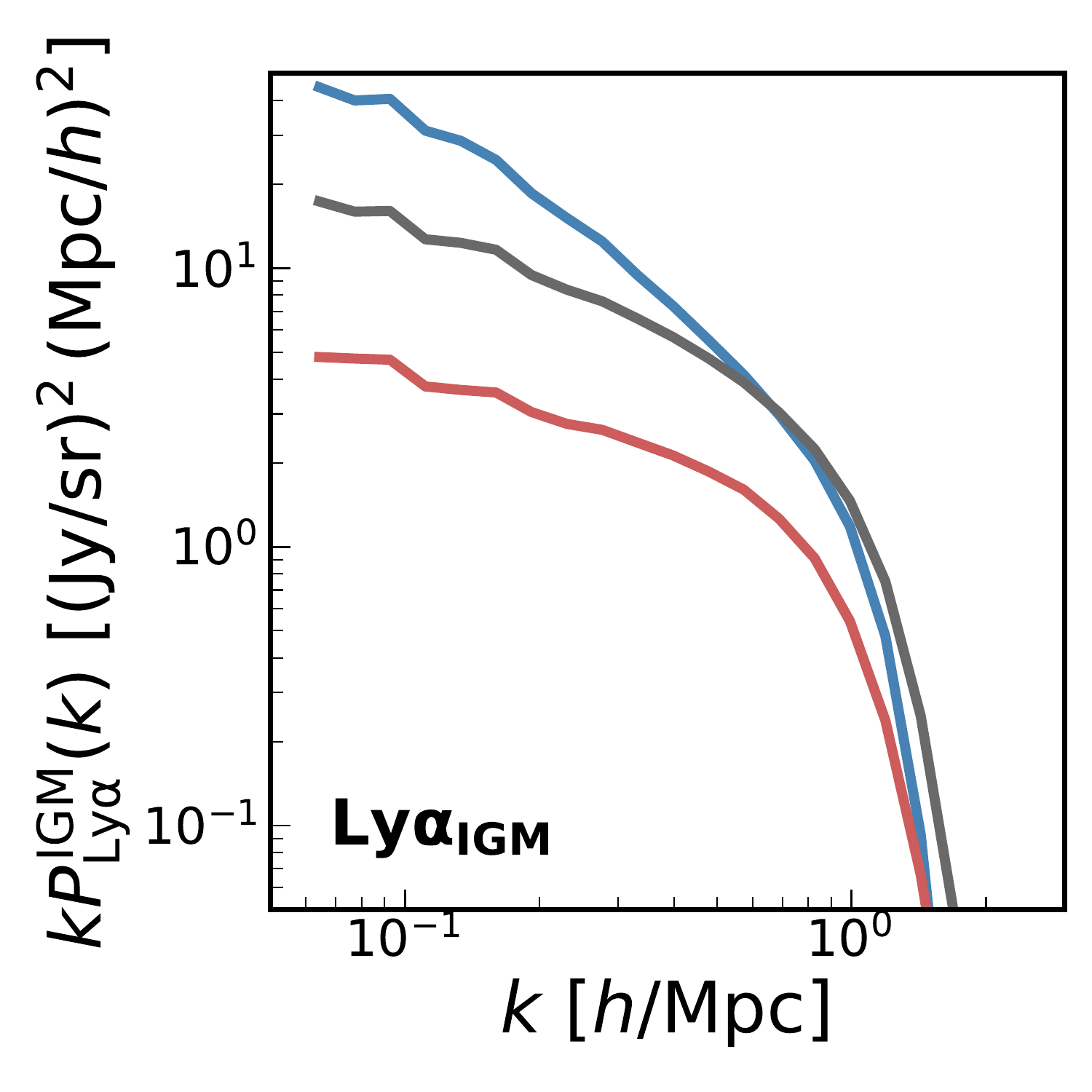} \includegraphics[width=0.24\textwidth]{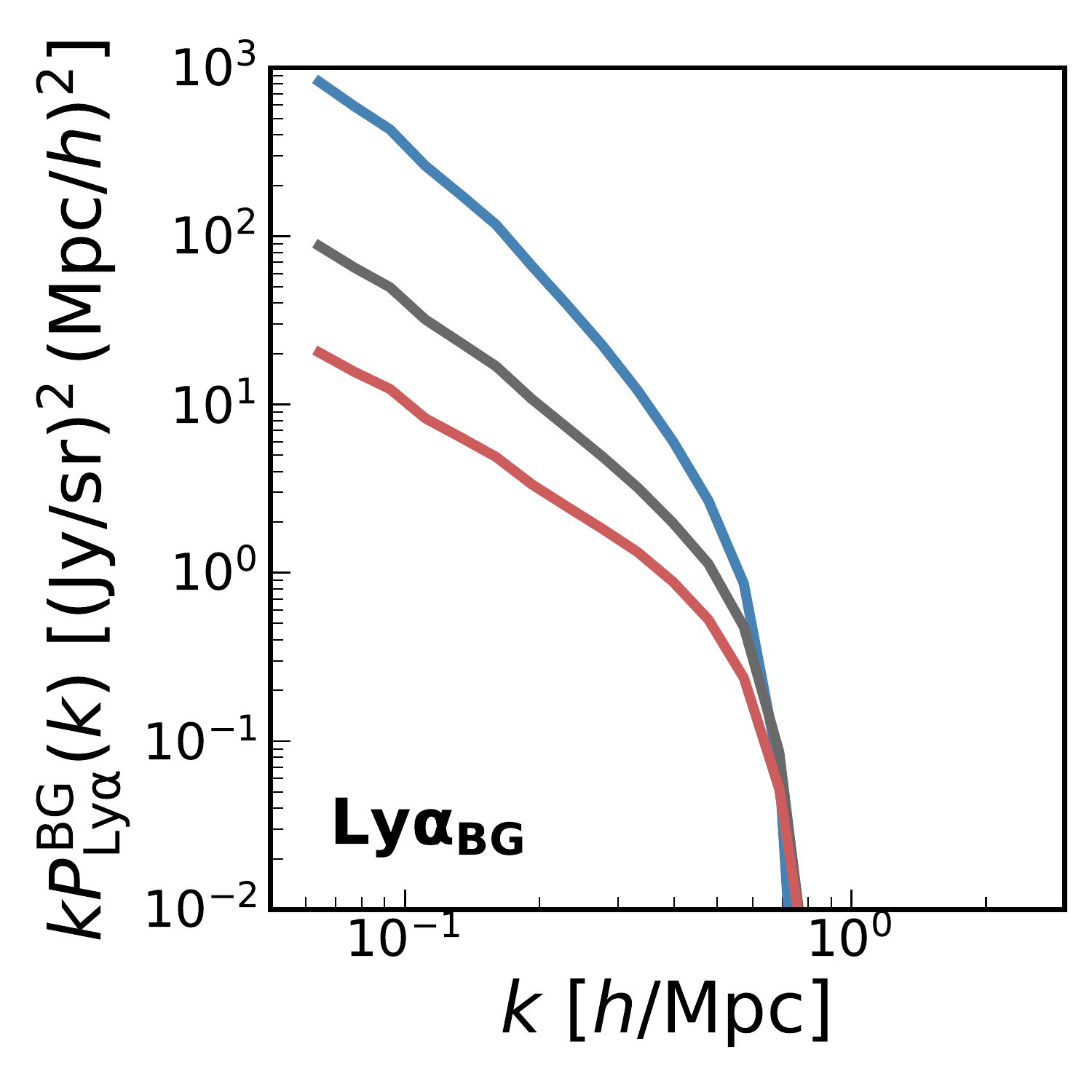}
 \includegraphics[width=0.24\textwidth]{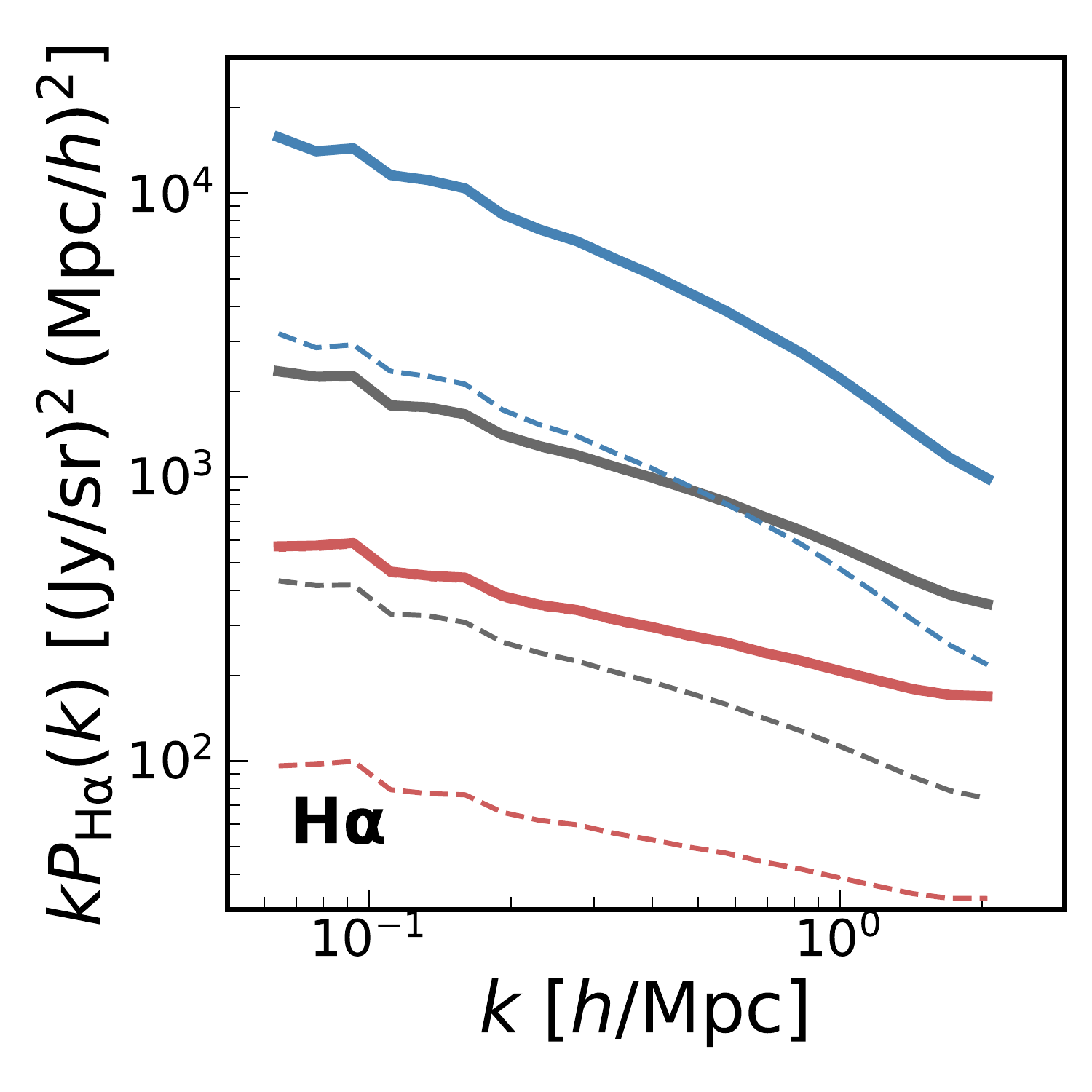} \includegraphics[width=0.24\textwidth]{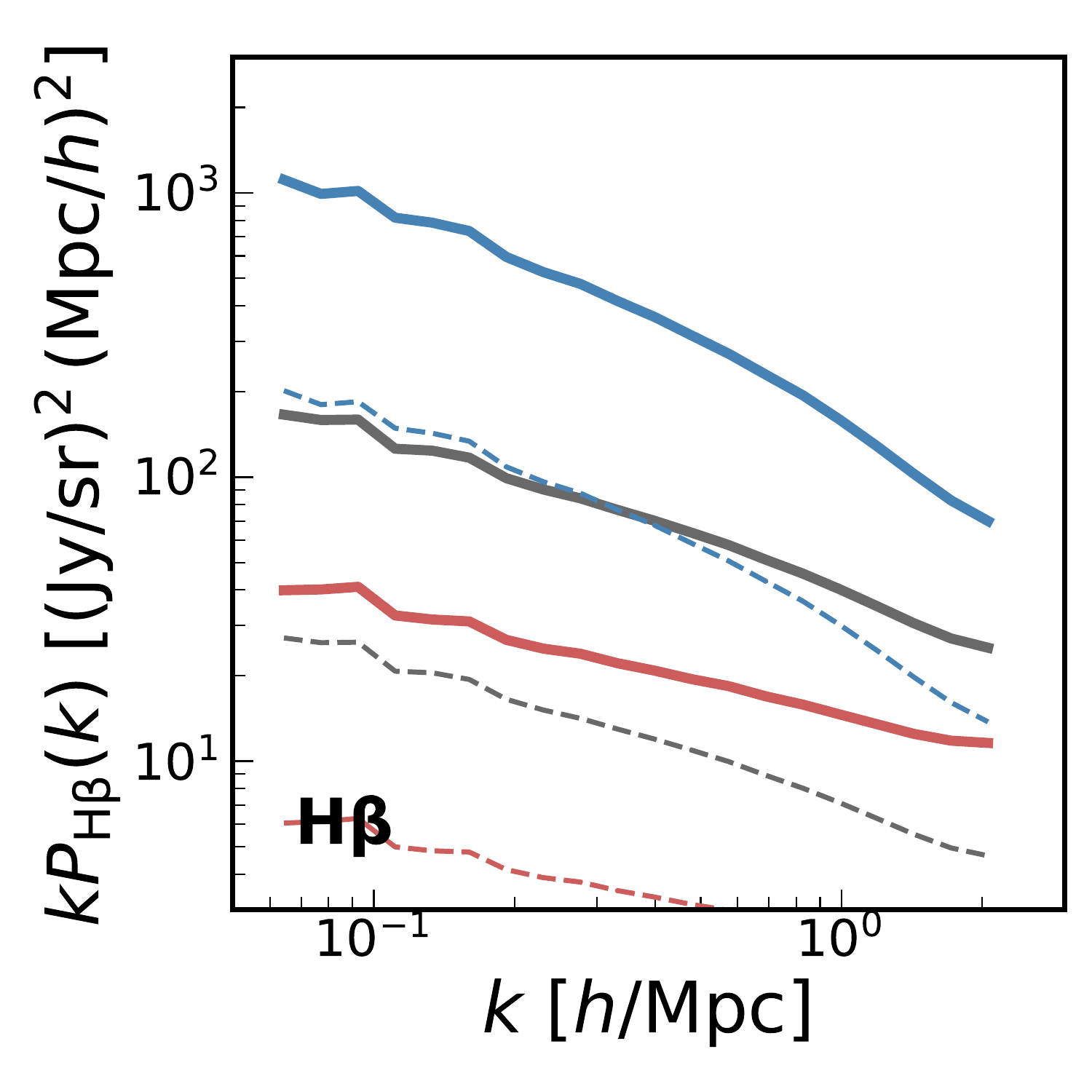}
 \includegraphics[width=0.24\textwidth]{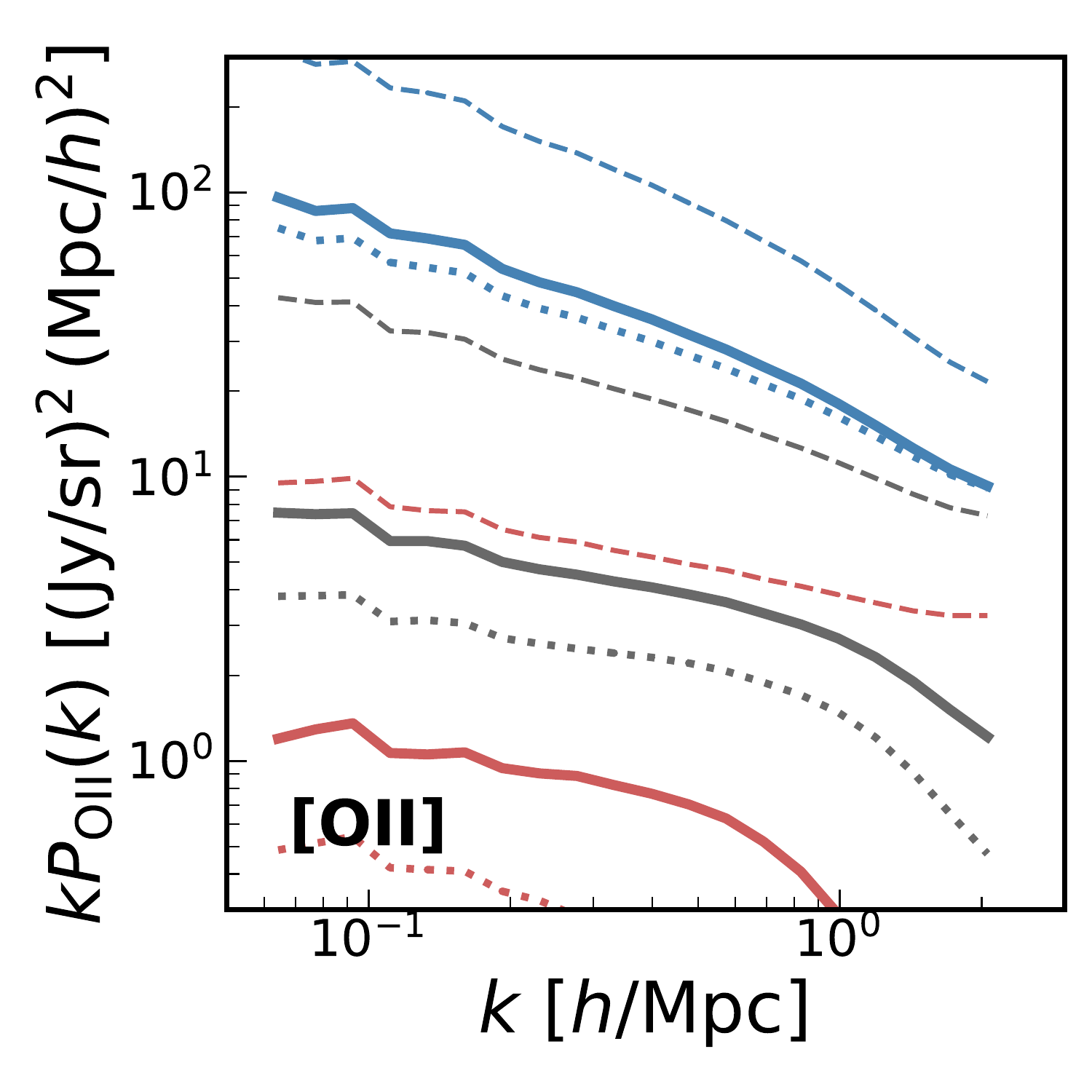} \includegraphics[width=0.24\textwidth]{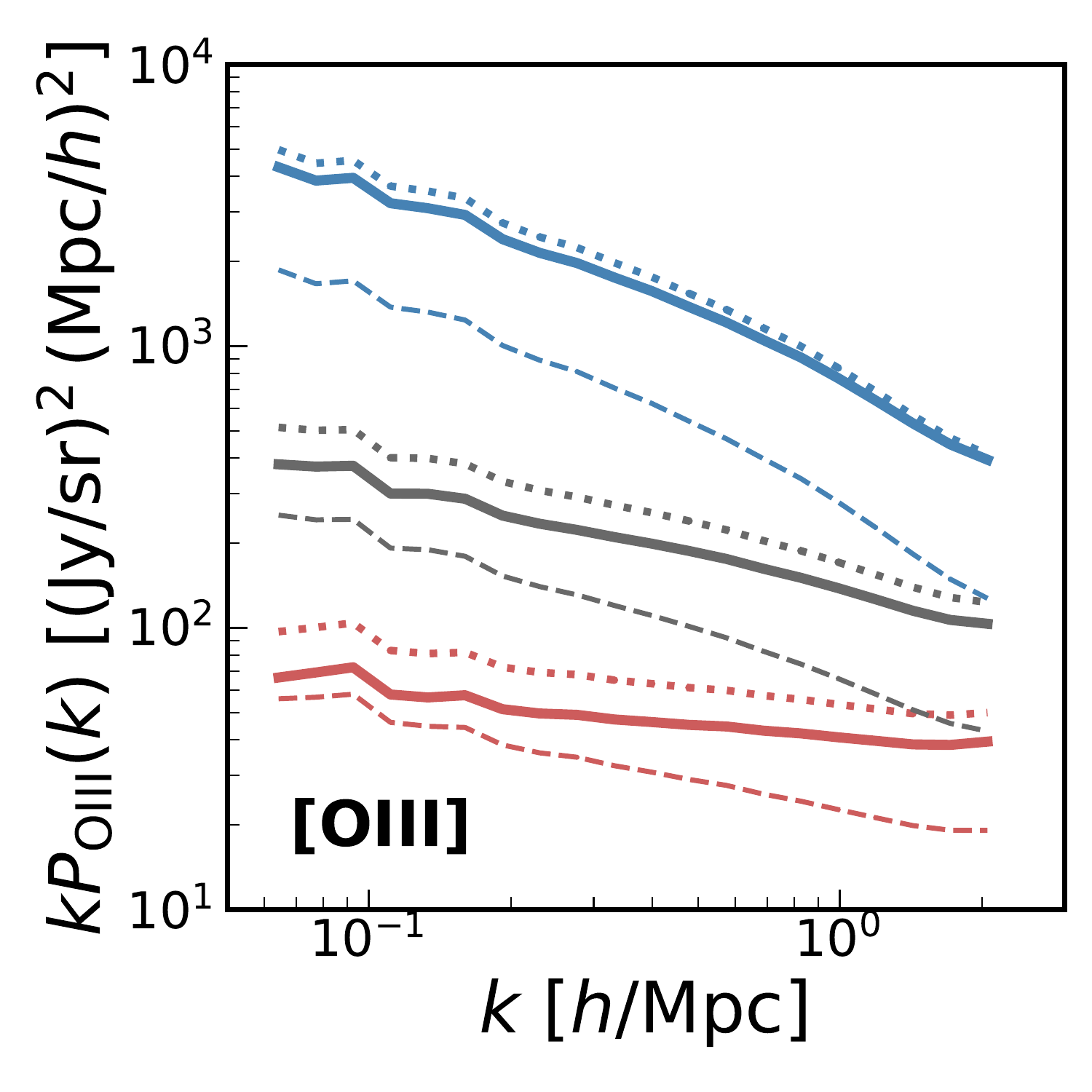}
 \caption{Power spectra of emission lines at $z=7$, 10, and 12, derived from the fiducial LIMFAST simulations (solid curves), together with two alternative cases assuming local scaling relations (dashed curves) and the anti-correlation $U \propto Z^{-0.8}$ (dotted curves) for the Balmer and oxygen lines. The RSD effect on the redshift-space power spectrum is displayed for the 21 cm line only by the dash-dotted curves.}
 \label{fig:ps}
\end{figure*}

Figure~\ref{fig:ps} shows the spherically-averaged power spectra of intrinsic, dust-unattenuated line emission at redshifts $z=7$, $z=10$, $z=12$, in the form of $kP(k)$. Results assuming local luminosity--SFR relations ($U \propto Z^{-0.8}$) are again shown by the dashed (dotted) curves for comparison. At all redshifts, clear differences exist between the shapes of power spectra for lines tracing star formation and the IGM. In particular, Ly$\alpha$ emission from the IGM shows a clear deficit of fluctuation power on small scales ($k \gtrsim 0.5\,h/\mathrm{Mpc}$), especially for the scattering background radiation which encodes the large Ly$\alpha$ ``horizon distance'' that smooths out small-scale fluctuations (note the many orders of magnitude shown in the $y$-axis of the Ly$\alpha_\mathrm{BG}$ panel; see also Figure~\ref{fig:lightcones}). The emission of Ly$\alpha$ (as well as other star-formation lines) from star-forming galaxies, on the other hand, has an overall flatter power spectrum, whose shape reflects how biased the line is as a LSS tracer. Compared with hydrogen lines, oxygen lines show stronger shape and amplitude evolution over $7<z<12$ due to the strong metallicity dependence of their luminosities. The (different) ways [\ion{O}{2}] and [\ion{O}{3}] depend on $U$ can also impact their power spectra differently, which may be used to probe ionization conditions of the ISM (see Section~\ref{sec:discuss:uvsz}). On small scales ($k\gtrsim1\,h/\mathrm{Mpc}$) and high redshifts ($z\gtrsim10$), in particular, the shapes of [\ion{O}{2}] and [\ion{O}{3}] power spectra deviate, with [\ion{O}{2}] being steeper and [\ion{O}{3}] being more flattened out, which can be seen from the comparison with the case assuming the local scaling relation where the line luminosity is simply proportional to the SFR (dashed lines). Such a systematic difference in the power spectrum shape can be attributed to the fact that [\ion{O}{2}] has a steeper luminosity--halo mass relation than [\ion{O}{3}], which gives rise to the (relatively) higher large-scale clustering power for [\ion{O}{2}].

Insofar as the model variations are considered, we see that orders of magnitude difference in the power spectrum amplitude can easily arise from seemingly modest changes of model assumptions like applying locally-calibrated scaling relations to high redshifts. Understanding such model dependence has important implications for making sensitivity forecasts and survey strategies of future LIM experiments, especially pathfinder programs that are usually instrument-noise-limited. A few additional examples of model variations and observational implications will be discussed in Section~\ref{sec:discuss:uvsz}. 

Finally, we note that considering the RSDs generally makes the spherically-averaged power spectra in real space and in redshift space only modestly different. We therefore only show a comparison for the 21 cm signals in Figure~\ref{fig:ps} (dash-dotted curves) for visual clarity. The net effect is a small increase in the fluctuation power on all scales, with the only exception of the 21 cm signal which instead shows a small decrease in power at $z=7$ due to the growing anti-correlation between the density and ionization fields. To better elucidate effects of the RSDs, we examine in Appendix~\ref{sec:psrsd} the 2D power spectra of a few representative line signals at $z=7$ and 11 (see Figure \ref{fig:psrsd}). Difference between real-space and redshift-space 2D power spectra is only discernible for the 21 cm and Ly$\alpha$ emission from the ionized IGM (without considering Ly$\alpha$ scattering effects), and is largely negligible for line signals tracing star formation that occurs in more overdense and thus more strongly biased regions. Interested readers are referred to Appendix~\ref{sec:psrsd} for a pedagogical overview of the implementation and impact of the RSDs in LIMFAST.

\section{Discussion}\label{sec:discussion} 

Below, we highlight the main features that distinguish LIMFAST from, and thus put our simulations into the context of, previous work from the literature, including the parent code 21cmFAST. Our goal here is to focus on model assumptions giving rise to qualitative differences that can be observationally tested (at least in principle) by various LIM signals, rather than scrutinizing quantitatively the likely causes of specific discrepancies between LIMFAST and the literature. Along these lines, we also discuss a few planned applications of LIMFAST in studies of the high-redshift universe exploiting the intensity mapping concept. 

\subsection{Improvements with Respect to 21cmFAST}\label{sec:vs21cmfast}

Besides an extension of the original 21cmFAST code to the more general regime of LIM, LIMFAST also has a few noteworthy improvements on simulations of the metagalactic and 21 cm radiation backgrounds, including the implementation of (1) a more physically-grounded galaxy formation model, and (2) redshift-dependent galaxy SEDs consistent with the cosmic chemical enrichment history implied by the galaxy formation model. While these changes made do not hugely impact the predicted 21 cm signals, understanding possible systematic effects caused by them is critical for both high-accuracy forward modeling and constraining astrophysics and cosmology with the simulated observables at cosmic dawn and reionization eras.

In terms of the galaxy formation model, we note that in 21cmFAST, for fast simulation of the ionization and spin temperature fields as key inputs to the 21 cm signal, the rate of star formation in dark matter halos is simply described by an efficiency parameter $f_*$ as a power-law of the halo mass and a star formation timescale given by $t_*/H(z)$, where $t_*$ takes a fixed value between 0 and 1. While these assumptions give a reasonable description of the co-evolution of galaxies and the IGM during the EoR, in detail some non-trivial discrepancies may occur, especially when the mass growth histories of individual halos are considered. 

\begin{figure}
 \centering
 \includegraphics[width=0.48\textwidth]{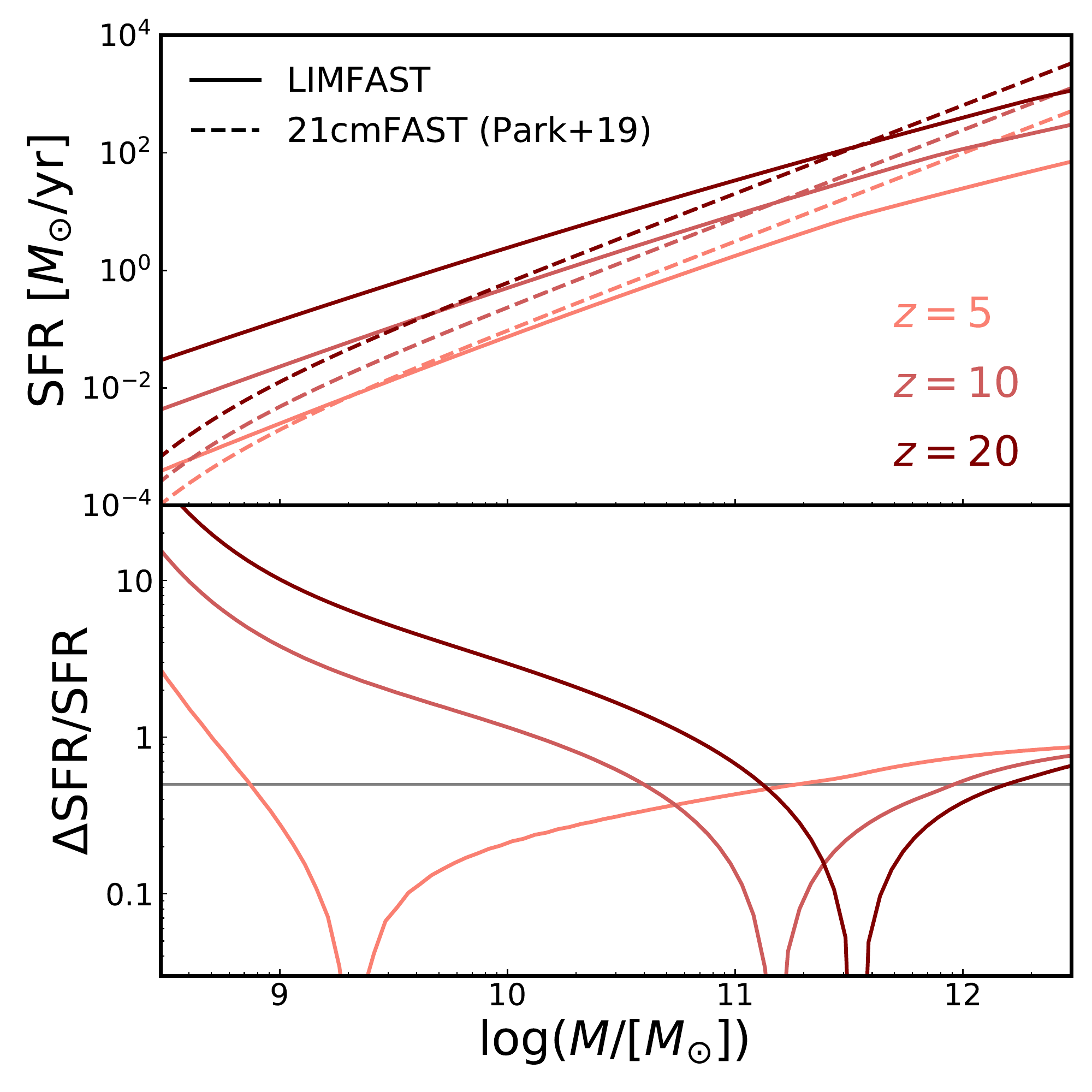}
 \caption{Top: comparison of the halo mass and redshift dependence of the SFR in LIMFAST (solid) and 21cmFAST (dashed), with the latter assuming best-fit parameters from \citet{Park2019} constrained by the UVLF, CMB, and quasar absorption spectra. Bottom: the fractional difference in the SFR from galaxy formation models adopted by LIMFAST and 21cmFAST. The horizontal line marks where the difference is 50\% of the SFR predicted by the \citet{Park2019} model.}
 \label{fig:comp_sfr}
\end{figure}

To illustrate this, in Figure~\ref{fig:comp_sfr} we compare the SFR as a function of halo mass and redshift given by the galaxy formation models adopted in LIMFAST and 21cmFAST, respectively. Note that for the model parameters in 21cmFAST, we take the best-fit values in Table~2 of \citet{Park2019} derived from observations of the UVLF, CMB optical depth, and quasar absorption spectra, instead of fiducial values reported there, which overall predict 10\% lower SFRs because they are not calibrated against observations. From the comparison, it is clear that even though both ways of parameterizing the SFR fit the observed UVLFs at $5 \lesssim z \lesssim 8$ reasonably well, the fiducial model adopted in LIMFAST implies a weaker mass dependence and a stronger redshift dependence of the SFR--halo mass relation, thus resulting in a more gradual SFRD evolution when integrated over the halo mass function as shown in Figure~\ref{fig:sfrdz}. The difference in the mass dependence is mainly attributed to the different ways feedback on star formation in low-mass halos is considered in LIMFAST and 21cmFAST. In LIMFAST, the SFR--halo mass relation is derived physically from star formation sustained by continuous accretion of baryonic matter onto dark matter halos and regulated by stellar feedback, of which the implied observational consequences are then checked against existing data. In 21cmFAST, the mass dependence and timescale of star formation is tuned to match the observational data without a clear physical interpretation in the context of feedback. Consequently, discrepant SFR--halo mass relations arise and lead to the difference in the SFRD that grows larger when extrapolated to higher redshifts. Different halo mass dependence of the SFR also leads to testable differences in the history and morphology of the reionization process tracked by statistics such as the bubble size distribution (BSD), as demonstrated in \citetalias{Sun_2022P2}. These differences underscore the impact of model dependence in simulating the onset of galaxy formation and its impact on the early phase of the EoR, especially in semi-numerical simulations where simple galaxy formation models are employed. 

The updated galaxy formation model implemented in LIMFAST also makes possible simulations of the metagalactic background consistent with histories of star formation and metal enrichment. Particularly, we can trace how SEDs of the radiation sources evolve over time as galaxies build up their stellar mass and become more metal-enriched. As described in Section~\ref{sec:code}, the fixed-value SED template used in 21cmFAST is replaced by the metallicity-dependent SED templates coupled with the metal production and enrichment history from our galaxy formation model. The impact of such redshift evolution of the source SEDs on the 21 cm signal can then be examined. 

\begin{figure}
 \centering
 \includegraphics[width=0.47\textwidth]{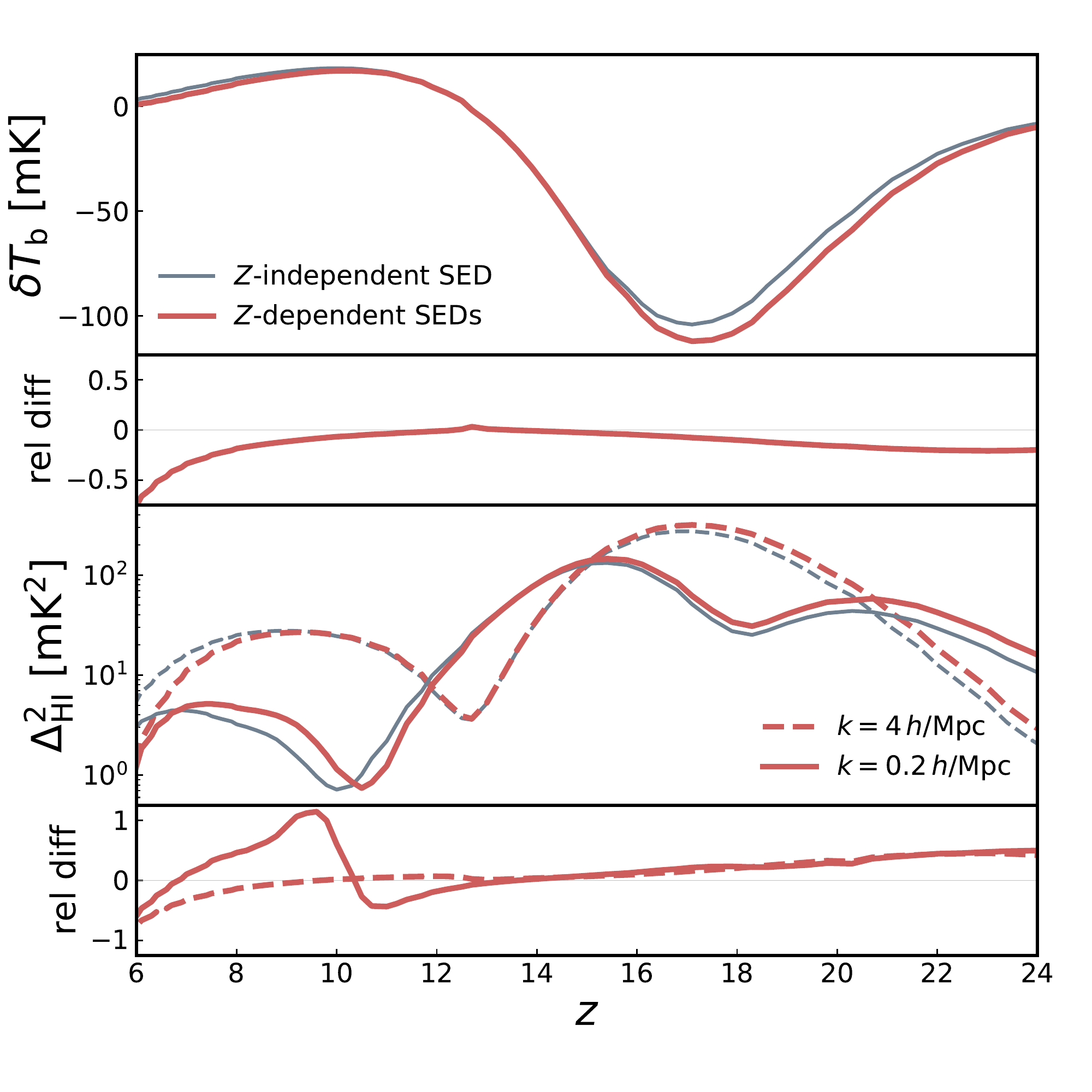}
 \caption{Effects of evolving galaxy SEDs on the 21 cm sky-averaged signal (top) and fluctuations (bottom). The same underlying galaxy formation model is assumed for both cases with and without invoking metallicity-dependent SEDs. Also shown in both panels is the relative difference defined as the \textit{fractional} deviation of the red curve from the gray curve.}
 \label{fig:comp_21cm}
\end{figure}

Figure~\ref{fig:comp_21cm} compares the 21 cm sky-averaged signal and the power spectrum as a function of redshift at a fixed scale ($k=0.2\,h/\mathrm{Mpc}$), predicted by our simulations with and without invoking metallicity-dependent SEDs. While the overall impact from metallicity-driven SED evolution is modest, clear differences in the timings of Ly$\alpha$ coupling ($z\sim20$) and reionization ($z\sim8$) can be seen, whereas the timing of the epoch of heating (EoH) driven by X-ray binaries ($z\sim15$) is little affected because of the metallicity-independent X-ray luminosity assumed. As a result, the asymmetries of the 21 cm sky-averaged signal is modified. These observations are consistent with how we expect the SED evolution with metallicity to modulate the 21 cm signals --- the overall harder spectrum resulted from an evolving, metallicity-dependent SED shifts the Ly$\alpha$ coupling and reionization epochs to higher redshifts. It also slightly reduces the contrast of fluctuations on large versus small scales and speeds up its evolution during the EoR. These effects can potentially introduce significant systematic uncertainties to efforts on using the statistics of the 21 cm signal to probe astrophysics and cosmology, such as the presence of massive Pop~III stars \cite[e.g.,][]{Mirocha2018, Mebane2020}, effects of cosmic rays \cite[e.g.,][]{Bera2022}, and alternative dark matter models \cite[e.g.,][]{Jones2021}.

Finally, we note that any metallicity dependence of the luminosity of X-ray binaries can similarly modulate the 21 cm signal during the EoH, whose implementation in LIMFAST is left for future work. In a recent study, \citet{Kaur_2022} find that different plausible prescriptions for the metallicity dependence of the X-ray luminosity correspond to order of unity changes in the 21 cm signals during the EoH. It is therefore interesting to understand, as a future step, how exactly the combined effect of metal enrichment on galaxy UV and X-ray spectra modifies the thermal and ionization histories of the IGM to be revealed by 21 cm and other complementary observations. 

\subsection{Comparison to Other Previous Work} \label{sec:discussion:lit}

In Section~\ref{sec:results}, we presented results from LIMFAST fiducial model and some simple variations, juxtaposed with a number of forecasts from previous LIM studies in the literature, without providing a careful explanation for the differences observed. Here, we outline a few factors that distinguish our results from the others, and identify the specific differences that they are responsible for. 

\subsubsection{High-Redshift Galaxy Formation}

As already pointed out in Sections~\ref{sec:sfrd} and \ref{sec:vs21cmfast}, the most significant difference in terms of high-$z$ galaxy formation between LIMFAST and other studies is that we predict overall higher and less steep SFRD as a function of redshift, especially at $z>8$ (see Figure~\ref{fig:sfrdz}). Two major factors embedded in our galaxy formation model together make the fractional contribution from faint galaxies to the total SFRD higher and thus are responsible for this discrepancy --- a small minimum halo mass, $M_\mathrm{min}$, for star formation that equals the atomic cooling threshold (assuming $T_\mathrm{vir}=10^4\,$K) and a relatively shallow, power-law relation between the amount of star formation and halo mass above that. Both factors modulate the SFRD evolution through the integral over the HMF, though $M_\mathrm{min}$ is relatively less important at the low-$z$ end considered, when a considerable fraction of halos have grown massive enough, e.g., $M > 10^{10}\,M_{\odot} \gg M_\mathrm{min}$. 

A useful example to consider is the SFRD from \citet{Silva2013}, which has a much steeper redshift dependence such that it predicts about 3 times more star formation at $z\sim6$ and about 5 times less at $z\sim10$. This is mainly because the \citet{Silva2013} model predates and thus is not calibrated against the much improved UVLFs measurements at $z>6$ \cite[e.g.,][]{Bouwens2015}, which put stringent constraints on the SFR--halo mass relation especially for faint galaxies. Even though star formation is allowed down to a similar halo mass cutoff of $10^{8}\,M_{\odot}$, \citet{Silva2013} infer a steeper SFR--halo mass relation that implies a UVLF faint-end slope now disfavored by observations. Indeed, for more recent, UVLF-calibrated models \cite[e.g.,][]{Munoz2022,Kannan_2022_LIM}, the predicted steepness of SFRD evolution is more comparable to that in LIMFAST, with the remaining discrepancy in amplitude largely due to the varying minimum mass cutoff assumed. 

Finally, we stress that the difference in the SFRD evolution, combined with the evolution it drives such as the chemical enrichment, serves as a major source of the discrepancies observed for the LIM signals, as will be discussed next. 

\subsubsection{LIM Signals at Cosmic Dawn and Reionization}

A fundamental distinction between the LIM signals predicted by LIMFAST and from the literature is that we have presented the intrinsic line emission, without accounting for any attenuation. The attenuation effect by intervening dust or neutral IGM can be treated on-the-fly or post-processed, analytically or numerically. Because of these different approaches and complications, and that the main goal of LIMFAST at this stage is to establish the connection between galaxy formation physics and the production of different emission lines, we leave the implementation of such attenuation effects to future work. Yet, it is still useful to compare the line signals at face value in terms of both their redshift evolution and spatial fluctuations measured by the power spectrum. 

For the 21 cm line, the fiducial case in LIMFAST predicts a global signal that reaches a minimum of depth $\sim-110\,$mK centered at $z\approx17$, which appears somewhat earlier in time compared with some recent studies involving similar assumptions about high-$z$ galaxy formation \citep[see e.g.,][where the trough locates at $z\sim12$]{Mirocha_2017,MF2019,Park2019}. The difference can be best understood by considering the fact that our fiducial case predicts an earlier onset of star formation, which presents a non-negligible amount of Ly$\alpha$ and X-ray backgrounds at $z>15$ that can modulate the 21 cm spin temperature. Indeed, models that produce similarly high SFRD by resorting to efficient star formation in low-mass halos or the presence of Pop~III stars (which we neglect in our fiducial simulations; but see \citealt{Parsons2022}) predict 21 cm signals centered at $z\gtrsim15$ as well \cite[e.g.,][]{Mesinger2016,Munoz2022}. At lower redshifts when reionization began, the 21 cm signal is increasingly driven by the ionization field, which is in turn sourced by the metagalactic ionizing radiation background that rapidly builds up as the SFRD rises. 

For Ly$\alpha$, Figure~\ref{fig:lines} shows good agreement between the shapes of redshift evolution from LIMFAST and \cite{Comaschi2016} for all three sources of Ly$\alpha$ emission considered, even though we purposely choose to leave out the dust correction in this work. This is likely because (1) both calculations assume the atomic cooling threshold for $M_\mathrm{min}$ and are calibrated against the observed UVLFs and CMB optical depth constraints, and (2) sources that contribute the majority of Ly$\alpha$ emission at these high redshifts are not strongly attenuated by dust, as is the case in \cite{Comaschi2016} at least. Generally, other studies like \cite{Pullen2014} and \cite{Heneka2017} show a steeper decrease towards high redshift with an overall lower amplitude because of the significant lower contribution from faint galaxies as discussed above, and the IGM attenuation in the Ly$\alpha$ damping wing may also play a role, especially when the neutral fraction is high \citep{Heneka2017}. On the other hand, for Ly$\alpha$ emission from recombinations in the diffuse ionized IGM, LIMFAST and \cite{Comaschi2016} both predict intensities a few to a hundred times lower than the previous findings by \citet{Pullen2014} and \citet{Heneka2017}. Because the global reionization history is roughly the same for all these studies, the offset in normalization might be caused by different model assumptions made for the recombination rate, such as the choice between Case A \citep{Heneka2017} vs. Case B (LIMFAST) recombinations  and homogeneous vs. inhomogeneous recombinations \citep{Park2019}. We also note that the Ly$\alpha$ emission due to IGM recombinations in \cite{Pullen2014} is derived from an SFRD that seems exceedingly high ($\dot{\rho}_* \sim 10\,M_{\odot}\,\mathrm{yr^{-1}\,Mpc^{-3}}$ at $z=6$). 

Similarly, for other star-formation lines like H$\alpha$ and [\ion{O}{2}], the redshift evolution in LIMFAST matches well that from THESAN~1 simulations, despite that small discrepancies exist for both the slope and the normalization as a result of differences in the cosmic star formation and photoionization modeling. There is not such a match for THESAN~2 because these simulations do not resolve the small halos that drive the signal at high redshift, resulting in a steeper evolution \citep[][]{Kannan_2022_LIM}. Finally, we emphasize again the subtle yet non-trivial differences in the redshift evolution and power spectrum arising from the usage or not of locally-calibrated scaling relations, and from different photoionization modeling assumptions (e.g., $\log U=-2$ vs. $U \propto Z^{-0.8}$) --- a useful feature that can be exploited to retrieve information about physical properties of the ISM \citep[][]{Silva2017}, which we will discuss in more detail next in Section~\ref{sec:discuss:uvsz}.

\subsubsection{Other Tools for Simulating Multi-tracer LIM}

Over the course of recent years, several other numerical tools have been developed for self-consistently simulating LIM signals of multiple tracers across the electromagnetic spectrum. These tools make use of either semi-analytic/empirical models of galaxy evolution to ``paint on'' galaxy properties (including emission lines) to the distribution of dark matter halos extracted from cosmological $N$-body simulations \cite[see e.g.,][]{Yang2021,Bethermin2022,SatoPolito_2022}, or fully cosmological magnetohydrodynamic (MHD) simulations to directly generate different line intensity fields of interest \cite[e.g.,][]{Kannan_2022_THESAN}. It is therefore interesting to compare them with LIMFAST, which has distinctive features of being semi-numerical, in order to understand the pros and cons of different tools available for the community. 

Compared to fully numerical simulations like THESAN \citep{Kannan_2022_THESAN}, the most notable difference of semi-analytic/empirical/numerical tools is the computational cost. Even though the former captures much more physical information about the evolution of radiation sources (e.g., details of gas dynamics and radiative transfer) in a highly integrated manner, compromises still need to be made due to the computational cost, which prevents the simulations from resolving the ISM-scale physics or reaching large, Gpc-scale volumes needed for probing astrophysics on large scales and cosmology. On the other hand, by running only $N$-body simulations without including baryonic physics, the ``paint-on'' method provides the access to larger cosmological volumes while still achieving the mass resolution necessary for properly capturing the statistics of halos and sub-halos. Depending on the type of questions that need to be addressed, prescriptions motivated by semi-analytic \cite[e.g.,][]{Yang2021,Bethermin2022} or semi-empirical \cite[e.g.,][]{SatoPolito_2022} galaxy evolution models may be adopted to paint light onto halos. 

The semi-numerical formalism LIMFAST inherits from 21cmFAST (Section~\ref{sec:code}) makes it even more computationally efficient to approximate the LSS formation and evolution as revealed by various LIM signals. The density-field-based method allows fast realizations (in a matter of hours) of multiple line intensity fields on a personal computer. While certain (potentially important) aspects of simulating the intensity fields and their synergies with other LSS tracers that are characteristic of ``paint-on" simulations have not been implemented in LIMFAST yet, such as the inclusion of scatter in various relations \citep{Reis2022,Murmu2023} and point sources like Lyman alpha emitters (LAEs), they can be included through relatively simple extensions of the current formalism using halo finders or subgrid sampling. 

Another important and related distinction to be considered is how complete and efficient the multi-scale and multi-phase cosmic gas may be simulated. None of the aforementioned ``paint-on'' simulations has so far incorporated calculations of the 21 cm signal during the EoR, although it is straightforward to post-process the density field realizations with methods like the excursion set formalism used in 21cmFAST/LIMFAST to describe the phase transition of the IGM and therefore the 21 cm signal. As demonstrated in \citetalias{Sun_2022P2}, capturing the complementarity between the 21 cm line and star-formation lines tracing opposite phases of the IGM during reionization plays a pivotal role in constraining the astrophysics of high-$z$ galaxy formation. The highly efficient (though approximate) simulations of structure formation and the production of multiple LIM signals that LIMFAST provides pave the way for fully Bayesian model inference studies in the near future, similar to what has been established for analyzing the 21 cm signal \citep{GM2015} to study both astrophysics and cosmology \citep{Park2019,Jones2021,Rudakovskyi2021,Munoz2022}. 

\subsection{Photoionization Modeling: Effects and Diagnosis} \label{sec:discuss:uvsz}

As introduced in Section~\ref{sec:sed}, while we have explored and generated grids of photoionization parameters like the metallicity $Z$ and ionization parameter $U$ that allow flexible modeling of the star-formation lines from \ion{H}{2} regions, it is unclear how exactly $U$ behaves among the entire source population and whether it connects to other galaxy properties like $Z$ at all. Therefore, potentially large uncertainties exist in our photoionization modeling, which substantially affect the predicted LIM signals, as has been demonstrated explicitly in Sections~\ref{sec:lumevo} and \ref{sec:ps}. In order to further illustrate this, we compare our fiducial case that assumes a fixed, arbitrarily-chosen value of $U$ with two ad hoc model variations that adopt physically-grounded but qualitatively different correlations between $U$ and $Z$. We further compare the predicted LIM signals of these models against the detectability of SPHEREx, in order to demonstrate how future LIM experiments might help improve the photoionization modeling of high-$z$ galaxies. 

The first case, which has already been briefly discussed and shown in Figures~\ref{fig:lines} and \ref{fig:ps}, builds on the wind-driven bubble model proposed by \citet{Dopita2006}, which predicts an anti-correlation $U \sim 3 U_c (Z/Z_{\odot})^{-0.8}$, where $U_c$ is a characteristic ionization parameter corresponding to $Z=0.1Z_{\odot}$. Qualitatively, the negative correlation in this case can be attributed to two effects --- stellar winds that become both more opaque and stronger at higher metallicities lead to more absorption of ionizing photons and more diluted ionizing flux due to larger shocked regions. The second case follows empirically the findings from photoionization models of nearby, MaNGA galaxies by \citet{JiYan2022}, which suggests $U \sim 0.15 U_c (Z/Z_{\odot})^{0.5}$. Such a positive correlation contradicts the expectation from the bubble model and might be a result of geometric effects of \ion{H}{2} regions in photoionization modeling (e.g., spherical vs. plane-parallel), although the exact cause of the correlation is still an open question that needs to be addressed by better data and models of the photoionization and gas dynamics in \ion{H}{2} regions. To be consistent with our fiducial choice of $\log U=-2$, we normalize both $U$--$Z$ relations such that they return $U=U_c=0.01$ for $Z=0.1\,Z_{\odot}$. 

Effects of combining $U$ and $Z$ differently in photoionization models can be conveniently projected into the observable space using, e.g., the O32--R23 diagram. The ionization-/excitation-sensitive line index $\mathrm{O32}=$ [\ion{O}{3}]/[\ion{O}{2}] and the oxygen-abundance-sensitive line index $\mathrm{R23}=$ ([\ion{O}{3}]+[\ion{O}{2}])/$\mathrm{H\beta}$ are often combined to jointly constrain the photoionization model in the analysis of individual \ion{H}{2} regions or galaxies \cite[e.g.,][]{Sanders2016,Strom2017}. Even though LIM observations do not resolve individual \ion{H}{2} regions or galaxies, it is helpful to understand how counterparts of these line indices, in terms of the mean line intensities directly probed by LIM data, compare under different assumptions of $Z$ and $U$. Both forward modeling and the interpretation of the mean line ratios as key observables in LIM data analysis benefit from such close-up views of the photoionization model. 

\begin{figure}
 \centering
 \includegraphics[width=0.48\textwidth]{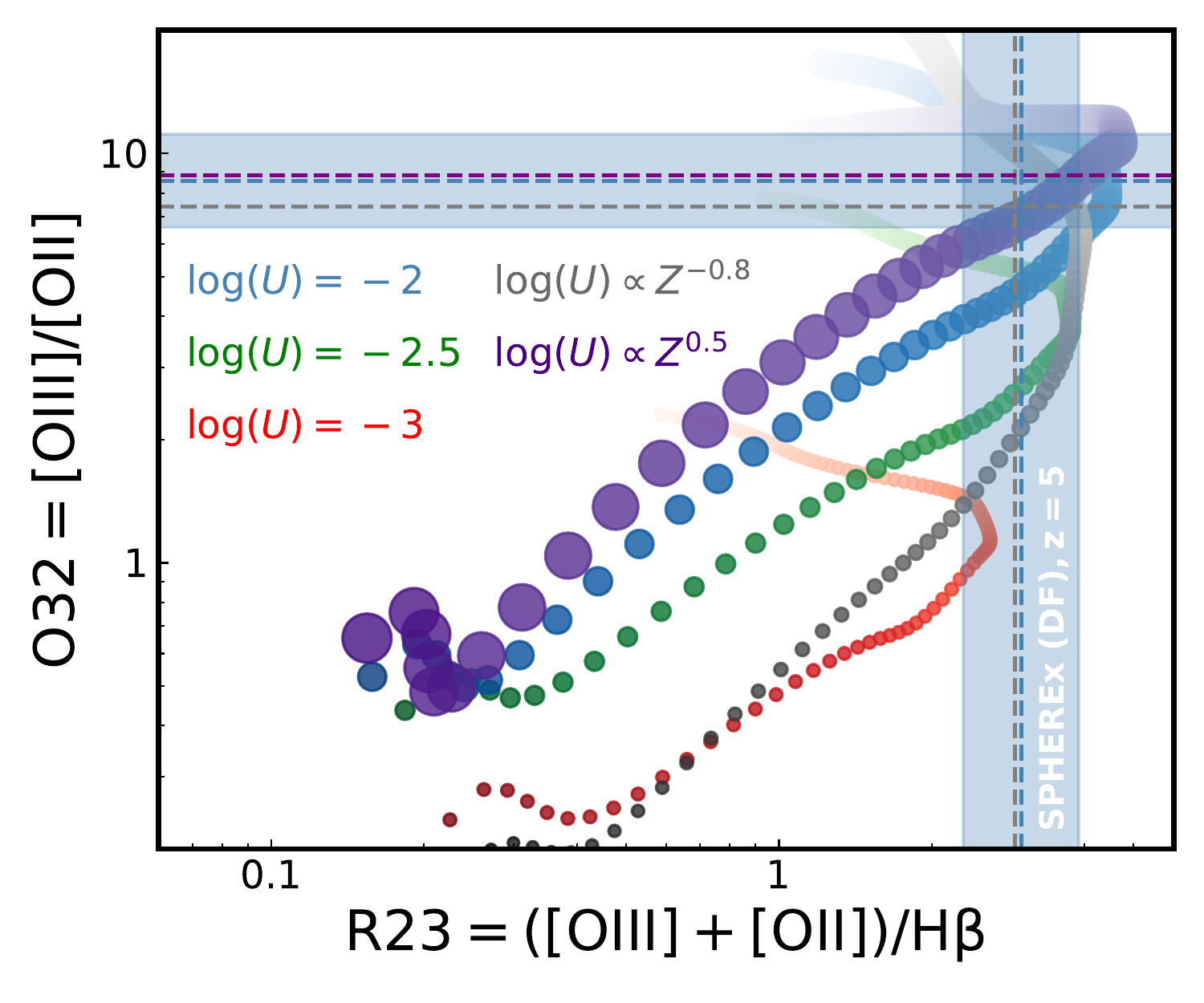}
 \caption{The relationship between O32 and R23 as a diagnostic of the photoionization model parameters $Z$ and $U$. Colored sequences of dots show O32 and R23 values of individual emitters with different combinations of $Z$ and $U$. The sequences with fixed marker size in blue, green, and red represent cases assuming $\log U=-2$ (fiducial), $-2.5$, and $-3$, with metallicities in range $-1.5<\log(Z/Z_{\odot})<0.5$ being indicated by the slowly evolving shades. For comparison, model variations assuming anti-correlated (correlated) $U$ and $Z$ are shown by the sequences in gray (purple), where $U$ decreases (increases) with $Z$ as indicated by the changing marker size \citep{Dopita2006, JiYan2022}, with bigger circles representing larger $U$. Dashed lines show O32 and R23 values computed from the mean intensities of individual lines at $z=5$, as simulated by LIMFAST in three different cases ($U=0.01$, $U \propto Z^{-0.8}$, and $U \propto Z^{0.5}$). The horizontal and vertical shaded bands shown for the fiducial case ($U=0.01$) indicate the projected constraints from the SPHEREx deep-field survey of the lines at $z=5$. }
 \label{fig:uvsz}
\end{figure}

Figure~\ref{fig:uvsz} shows how O32 and R23 respond to changes in $U$ and its possible dependence on $Z$. Predictions from five cases of the $U$--$Z$ relation are shown by the sets of points in different colors, including the two aforementioned model variations based on findings of \citet{Dopita2006} and \citet{JiYan2022}. Cases with fixed $U$ show similar trends for O32 versus R23, with the normalization primarily set by the exact $U$ value. The non-monotonic and double-valued behavior at high R23 and O32 is caused by the $Z$ dependence of the oxygen line strengths, which increase and decrease with $Z$ at the low-$Z$ and high-$Z$ ends, respectively (see also Figure~\ref{fig:udependence}). The same general behavior persists for cases with correlated $U$ and $Z$, despite that when $U$ and $Z$ are anti-correlated the metallicity effect is partially counteracted by the corresponding $U$ evolution, making the turnover at intermediate $Z$ much more gradual. 

On top of pairs of O32 and R23 shown by the scattered points, we also overplot the corresponding line indices formed by the derived mean line intensities $\bar{I}_\mathrm{[O\,II]}$, $\bar{I}_\mathrm{[O\,III]}$, and $\bar{I}_\mathrm{H\beta}$ at $z=5$ (the lowest redshift covered by current LIMFAST simulations), as shown by the horizontal and vertical dashed lines for cases with $U=0.01$, $U \propto Z^{-0.8}$, and $U \propto Z^{0.5}$. While these quantities cannot be directly mapped to the mean physical conditions of photoionization in $z=5$ galaxies due to its highly heterogeneous nature among the galaxy population, measuring and comparing them with predictions from the photoionization model still provides useful insights into the permitted parameter space. With the shaded bands, we illustrate the levels of uncertainty that can be achieved by the 200\,deg$^2$ SPHEREx deep-field survey for these line indices. Assuming the nominal, deep-field survey depth\footnote{See the publicly-available SPHEREx sensitivity data product \url{https://github.com/SPHEREx/Public-products/blob/master/Surface\_Brightness\_v28\_base\_cbe.txt.}}, we expect SPHEREx to measure both O32 and R23 with $\mathrm{S/N} \approx 3$ at $z=5$. Clearly, extreme scenarios that deviate significantly from our fiducial case, e.g., $\log U = -3$ (red), can be readily tested with the LIM-derived O32 and R23, but it is challenging for LIM measurements to distinguish more similar cases like those differing only in how $U$ scales with $Z$ at high redshift. Given how complicated the detailed picture of photoionization might become and vary for different galaxies, reliably forward modeling or inferring from the LIM signals of lines such as [\ion{O}{2}] and [\ion{O}{3}] can be quite challenging, especially at high redshifts where prior knowledge on the physical conditions of \ion{H}{2} regions is limited. Imminent JWST observations of these lines in individual high-$z$ galaxies will therefore serve as a resourceful guide for how to better model and interpret their LIM signals in the near future \cite[e.g.,][]{Katz_2023}. 

\subsection{Applications of LIMFAST}
The density field-based nature of LIMFAST simulations makes them suitable for efficiently generating many realizations in the forward modeling and analysis pipeline, which is pivotal to the success of any intensity mapping experiment. As has been demonstrated by 21CMMC \citep{GM2015} for analyzing the 21 cm signal, it is also possible to build up a full Bayesian inference framework for parameter studies, provided that a flexible, generative model can be constructed, which is what we aim to demonstrate with LIMFAST for investigations of the multi-tracer LIM concept. A few plans have been made so far to implement LIMFAST into the simulation pipeline of forthcoming intensity mapping surveys, including the Tomographic Ionized-carbon Mapping Experiment (TIME; \citealt{Crites2014,Sun2021}) and the Spectro-Photometer for the History of the Universe, Epoch of Reionization, and Ices Explorer (SPHEREx; \citealt{Dore2014}). While LIMFAST only runs down to $z=5$, the lack of correlation between large-scale structure at distinct redshifts allows us to combine with simulations dedicated to the post-reionization universe ($0<z<5$) and create complete light cones that include both the target signal and different sources of contamination. 

Meanwhile, the modular nature of LIMFAST makes it convenient to implement model variations and theoretically study their effects on various observables. Examples of this include the different forms of stellar feedback and the star formation law investigated in \citetalias{Sun_2022P2}, and the impact of Pop~III star formation with different IMF assumptions on the LIM signals of \ion{He}{2} and H$\alpha$ at cosmic dawn explored in \citet{Parsons2022}. We have also extended our metagalactic radiation background model to simulate intensity mapping observations of the continuum emission at near-infrared, where the  cosmic near-infrared background reveals rich information about the collective properties of the first stars and galaxies as well as their impact on reionization \citep{Sun2021NIRB}, especially when jointly analyzed with the 21 cm signal (Sun et al. in prep). 

\section{Conclusions}\label{sec:conclusion} 

In this paper, we have introduced LIMFAST, a semi-numerical code to physically model the production of emission lines in galaxies and the IGM during the cosmic dawn and reionization eras and simulate their LIM signals. LIMFAST implements the analytic galaxy formation and evolution models by \cite{Furlanetto_2021} into the framework of structure formation and reionization simulations in 21cmFAST, and utilizes pre-computed look-up tables from stellar population synthesis (BPASS) and photoionization modeling (\textsc{cloudy}) to obtain emissivities of the lines of interest for LIM studies. 

We have detailed the galaxy formation model, the calculations of stellar and nebular SEDs, and the various kinds of line emission that can be simulated (Section \ref{sec:code}). We have showcased physical evolution of the galaxy population and the IGM derived from LIMFAST runs, followed by a collection of predicted LIM signals in terms of their light cones, mean line intensities, and power spectra. Emission line signals presented include the 21 cm line, the Ly$\alpha$ IGM and background radiation fields, as well as the Ly$\alpha$, H$\alpha$, H$\beta$, [\ion{O}{2}] $3727 \angs$, and [\ion{O}{3}] $5007 \angs$ lines arising from star formation (Section~\ref{sec:results}). We have compared our results with a number of previous studies from the literature including the parent code 21cmFAST, discussed key aspects that distinguish LIMFAST from them, and introduced some case studies enabled by LIMFAST (Section \ref{sec:discussion}). Several aspects of our simulations are noteworthy, which we summarize and reiterate as follows: 

\begin{enumerate}

\item With LIMFAST, we are able to generate histories of cosmic star formation, reionization, and metal enrichment in a fully self-consistent manner that is in good agreement with the observational constraints available so far. Due to some assumptions we make --- that star formation in galaxies is regulated by momentum-driven feedback from supernovae and stars can form in halos of mass as low as the atomic cooling limit with $T_\mathrm{vir}=10^4\,$K, we predict overall a higher rate of cosmic star formation at $z>6$ that evolves less strongly with redshift compared with most previous studies in the literature. We note, though, that this picture of high-$z$ galaxy formation still agrees well with observations and implies a plausible reionization history. Neglecting other complications that also modulate the line strengths, we argue that higher SFRD in general leads to stronger emission in the various star-formation lines considered in this work. 

\item Tracing how chemical enrichment drives the SED evolution of the galaxy population, as sources of both the ionizing radiation and the line intensity fields, is important for accurately modeling the LIM signals of interest. Even for the 21 cm line whose emissivity is not directly related to the galaxy SEDs, order of unity differences in both sky-averaged and fluctuation signals may result from ignoring the SED evolution. As can be seen from the comparison with cases assuming locally-calibrated relations, this has an even larger impact on the star-formation lines whose emissivities are usually strong functions of metallicity $Z$. This marks an important source of astrophysical uncertainty to be considered in future  modeling and analysis of LIM observations, especially for studying cosmology. 

\item Since we choose to not apply corrections for the attenuation effects on line intensities due to dust and neutral gas, which themselves are highly uncertain and model dependent, it is somewhat difficult to make fair comparisons between results from our simulations and those in the literature with corrections applied. Nonetheless, qualitative features like the steepness of redshift evolution help understand the source(s) of discrepancies. Overall, we find good qualitative agreement among predictions based on models calibrated to latest observational data including the UVLFs at $z>6$. 

\item Detailed photoionization (or more generally ISM) modeling is sophisticated and usually difficult to be done consistently with simplistic models of galaxy formation and evolution. Using simple assumptions and ad hoc variations, we have demonstrated challenges posed by this issue for modeling the line signals (e.g., [\ion{O}{2}] and [\ion{O}{3}]) and also showed that it is useful to consider combinations of the line ratios as a way to constrain key parameters in the photoionization modeling such as the ionization parameter $U$. By analogy to the strong line calibration method in the study of chemical evolution of individual galaxies, we have illustrated how $Z$ and $U$ may be jointed probed by observational data from LIM surveys like SPHEREx. 

\item Although LIMFAST in its current form is obviously too simplistic to be applied to all kinds of problems in the study of the high-redshift universe using the intensity mapping technique, interesting applications of it to real-world LIM efforts are already underway. Thanks to its modular nature (Figure~\ref{fig:flowchart}), we expect new features to be implemented and complement the existing framework as new astrophysical and cosmological applications being pursued. 

\end{enumerate}

To conclude, following up on this methodology paper, in \citetalias{Sun_2022P2} of this series we take into account models of the neutral ISM for the inclusion of the [\ion{C}{2}] 158 $\mu$m and CO lines, which are popular target lines for sub-mm LIM. We anticipate further implementations such as the aforementioned attenuation effects due to neutral gas and dust, which can be self-consistently included into the existing formalism. LIMFAST enables time-efficient multi-tracer simulations of large cosmological volumes, which make it a useful complement to more sophisticated numerical simulations such as the recently presented THESAN simulations \citep{Kannan_2022_THESAN}, for not only forward modeling but also model inference from the LIM data sets. 

\begin{acknowledgments}

We thank the anonymous reviewer for their helpful comments that improve this paper. We are grateful to Steven Furlanetto, Adam Trapp and Fred Davies for useful discussions and comments throughout this project, as well as Rahul Kannan and Yuxiang Qin for their valuable comments on the early draft of this paper. We acknowledge support from the JPL R\&TD strategic initiative grant on line intensity mapping. This research was carried out at the Jet Propulsion Laboratory, California Institute of Technology, under a contract with the National Aeronautics and Space Administration. 

\software{ARES \citep{Mirocha_2017}, BPASS \citep{Eldridge2017}, \textsc{cloudy} \citep{Ferland2017}, 21cmFAST \citep{Mesinger2011}}

\end{acknowledgments}

\appendix

\twocolumngrid

\begin{figure*}
 \centering
 \includegraphics[width=0.9\textwidth]{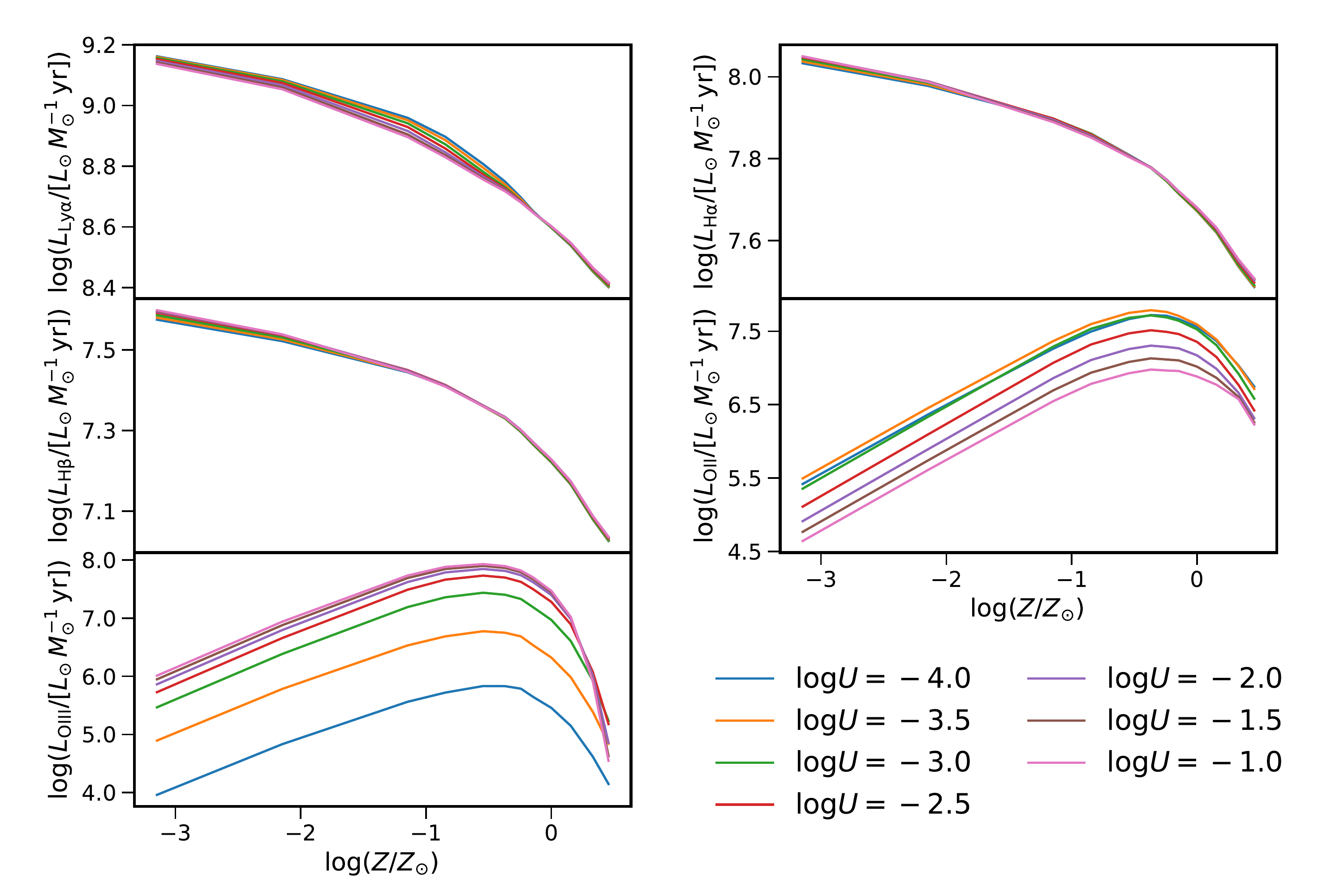}
 \caption{Dependence of the  luminosity with metallicity and ionization parameter for lines arising from star formation. Note the drastically different scales in the vertical axes for different lines.}
 \label{fig:udependence}
\end{figure*}

\section{Luminosity Dependence on $U$ and $Z$}\label{sec:udependence} 

Figure \ref{fig:udependence} shows the dependence of the luminosity of lines from star formation on metallicity and ionization parameter. The luminosity of the oxygen lines changes by orders of magnitude when varying the metallicity, while the changes in the hydrogen line luminosities are only of a factor of a few. Furthermore, varying the ionization parameter also has a major impact (up to an order of magnitude) on the oxygen lines, with opposite effects for the \ion{O}{2} and \ion{O}{3} ions, whereas the impact on the hydrogen lines is less than order of unity. 

\section{Redshift-Space Distortions: A Pedagogical Description}\label{sec:psrsd} 

We briefly revisit here the formalism of RSD, and show its application to LIM observations targeting 21 cm and other emission lines. Our discussion is based on that in \cite{Mao2012}, to which we refer the interested reader for details.

Observing an emission line at a specific frequency indicates, in principle, the position along the LOS of the sources producing the line. In reality, however, the observed redshift can be different from the real cosmological one because not only the Hubble expansion determines its value; the LOS peculiar velocity of the sources affects the value of the observed frequency via the Doppler effect which, in turn, results in deviations from the true redshift. Quantitatively, when peculiar velocities along the LOS, $v_\parallel$, are present, the position $\boldsymbol{r}$ of a source in real space appears to a position $\boldsymbol{s}$ in redshift space through the expression 
\begin{equation}
\boldsymbol{s} = \boldsymbol{r} + \frac{1 + z_{\rm obs}}{H(z_{\rm obs})} v_\parallel \hat{r} ~.
\end{equation}
Here, $1 + z_{\rm obs} = (1 + z)(1 - v_\parallel/c)^{-1}$, $\hat{r}$ is the unit vector along the LOS direction, and $z_{\rm obs}$ and $z$ are the observed and cosmological redshifts, respectively. Because the peculiar velocities of the sources are typically unknown in observations, the redshift space information is the only one that is observable. Therefore, to predict observed quantities from our simulations, we need to account for the effect of peculiar velocities on calculations in real space. We use the term redshift-space distortions, or RSDs, here to refer to the change in apparent position induced by the peculiar velocity of the sources \citep[see][for a review on RSDs]{Hamilton1998}. 

As detailed in Section 4 in \cite{Mao2012}, deriving the cosmological radiative transfer equations for radiation emitted from sources  
with peculiar velocity, and considering an optically-thin medium along the LOS, one finds that the observed intensity can be written as \citep[Equation 12 in][]{Mao2012} 
\begin{equation}
I_{\nu_{\rm obs}} = \frac{c L_{{\nu_0}} \,a\, \nu^2_{\rm obs}}{4 \pi \nu^3_0 H(a)} \frac{n^\prime }{\left|1 +  (dv_\parallel/dr_\parallel)/[aH(a)]\right|} ~.
\end{equation}
Here, $L_{{\nu_0}}$ is the rest-frame luminosity of the sources, $H(a)$ represents the Hubble parameter at 
scale factor $a$, and $ \nu_{0}$ and $ \nu_{\rm obs}$ are the rest-frame and observed frequencies of the emission, respectively. 
The rightmost term denotes the apparent change in the number density of sources due to their peculiar velocities, where $n^\prime $ is the \textit{proper} number density of sources in real space, and ${dv_\parallel}/{dr_\parallel}$ is the velocity gradient along the LOS. Expressing now the observed frequency in terms of the rest-frame one via $\nu_{\rm obs} = \nu_0 (1+ z)^{-1} (1 - v_\parallel/c)$ and considering the \textit{comoving} number density of sources, $n$, we can rewrite the previous equation as 
\begin{align}\label{eq:iobs}
I_{\nu_{\rm obs}} & = \frac{c L_{{\nu_0} } (1 - v_\parallel/c)^2 }{4 \pi \nu_0 H(z)} \frac{n}{\left|1 + (1+z) (dv_\parallel/dr_\parallel)/H(z)\right|}  ~  \nonumber \\
 & \simeq \frac{ I_\nu (z)}{\left|1 + (1+z) (dv_\parallel/dr_\parallel)/H(z)\right|} ~ .
\end{align} 
This highlights the fact that the observed intensity considering the RSDs is simply roughly equivalent to the specific intensity computed by LIMFAST (Equation \ref{eq:intens}), with a velocity gradient correction denoted by the denominator. Here, we have omitted the correction term $(1 - v_\parallel/c)^2$ in the numerator because its effect is much smaller than that from the denominator. The correction shown in the last term of Equation~(\ref{eq:iobs}) is the one that we will apply to the intensity of the optical and UV lines computed by LIMFAST. This is the same correction as for the 21 cm radiation field shown in Equation~(\ref{eq:tbeq}), but there the 21 cm signal is expressed as a differential brightness temperature instead of as an intensity. 

As mentioned before, the above derivation considers an optically-thin medium, which is valid for all optical and UV emission lines considered in our work except for Ly$\alpha$. The reason for this difference is that Ly$\alpha$ is severely affected by the neutral hydrogen gas, both within halos and in the IGM, due to its resonant nature \citep[see a review on the physics of Ly$\alpha$ in][]{Dijkstra2017}\footnote{See a discussion about the optically thin assumption and its validity for 21 cm emission in \cite{Mao2012}.}. For simplicity, we do not address further corrections for Ly$\alpha$ here, but we refer the interested reader to \cite{Zheng2011} for an investigation of the impact of Ly$\alpha$ radiative transfer to the RSDs. 

\begin{figure*}
 \centering
 \includegraphics[width=\textwidth]{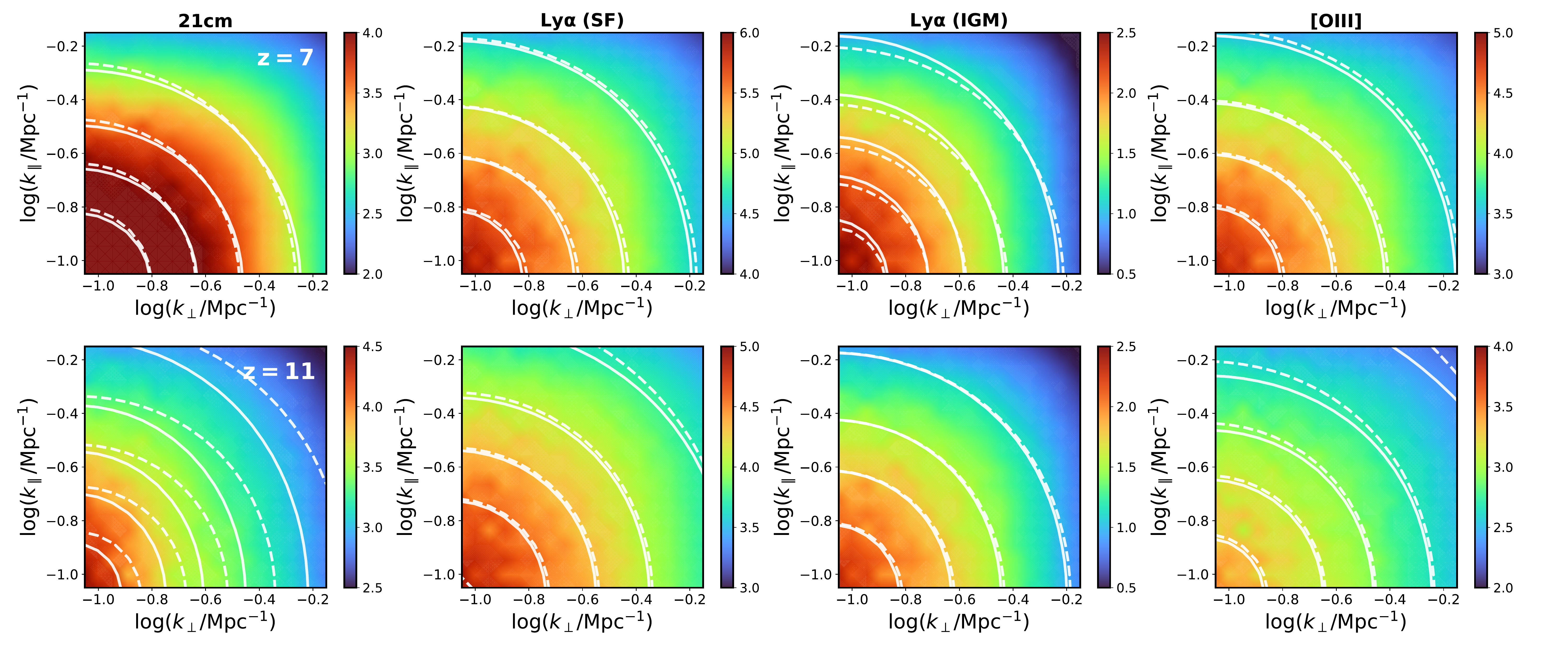}
 \caption{From left to right, 2D power spectra of the 21 cm, Ly$\alpha$ (from star formation and the ionized IGM), and [\ion{O}{3}] lines simulated by LIMFAST at $z=7$ (top row) and 11 (bottom row), where $\bar{x}_\mathrm{HI} \approx 0.3$ and 0.9, respectively. The overlaid contours highlight the shape difference of the power spectra, with the solid and dashed ones denoting cases with and without the RSD effect, respectively. The amplification/suppression and anisotropies induced by the RSDs vary for different lines at different redshifts.}
 \label{fig:psrsd}
\end{figure*}

Finally, the observable of interest that is sensitive to the RSDs is the power spectrum. Thus, it is also useful to write the intensity (or brightness temperature) in terms of fluctuations to visualize the impact from the clustering bias of the sources and from the RSDs when comparing the power spectra of different emission lines. For the intensity case and considering linear scales, 
\begin{equation}\label{eq:fluctuations}
I_{\nu_{\rm obs}} = \bar{I}_\nu(z)\, \frac{1 + b\,\delta_{\rm DM} (\bf{r})}{|1 + \delta_{\partial_r v} ({\bf r})|}.
\end{equation}
Here, $\bar{I}_\nu(z)$ is the cosmic mean intensity of a given emission line, $\delta_{\rm DM} (\bf{r})$ denotes the fluctuations of the dark matter density field, and $\delta_{\partial_r v} ({\bf r}) = [(1+z)/H(z)] dv_\parallel/dr_\parallel$, with all three quantities considered in real space. The term $b$ denotes the bias factor of the line emission with respect to the dark matter field, where we assume this emission bias to be the same as that of the sources since we ignore radiative transfer effects. Overall, Equation~(\ref{eq:fluctuations}) readily shows the known result that the impact of the RSDs will be more significant when the sources of radiation are less biased. This case corresponds to emission sourced by the intergalactic gas, i.e., 21 cm and IGM Ly$\alpha$\footnote{We remind here that we do not account for the Ly$\alpha$ radiative transfer that can also affect the fluctuations of this emission line.}, while emission from star formation tracing much more biased galaxies will be less sensitive to the RSDs. 

Figure \ref{fig:psrsd} displays the two dimensional power spectra in redshift space for a selection of lines simulated by LIMFAST at redshifts $z=7$ (top row) and $z=11$ (bottom row). Each panel is color coded by the power, i.e., $I_\nu^2 / V$, in units of $\mathrm{mK^2\,Mpc^3}$ for the 21 cm line or $\mathrm{(Jy/sr)^2\,Mpc^3}$ otherwise. The overlaid contours contrast the shape difference between real-space (dashed) and redshift-space (solid) power spectra, to highlight the anisotropies in redshift space induced by the RSDs. Overall, the RSDs are most visible for lines that trace and are sourced by the intergalactic gas, i.e., the 21 cm line and the intergalactic Ly$\alpha$ radiation. When the clustering bias of the sources is large, as is the case for star-formation lines, the impact of the RSDs is less important, as indicated by Equation~(\ref{eq:fluctuations}). Also noteworthy is that the RSDs appear to affect the 21 cm 2D power spectra in opposite ways at $z=7$ and 11, which correspond to IGM neutral fractions $\bar{x}_\mathrm{HI} \approx 0.3$ and 0.9, respectively. This is mainly due to the progress of the reionization process, when an anti-correlation between the ionization and density fields contributes more to the 21 cm fluctuations as the IGM becomes more ionized \cite[see the discussion in][for more details]{Jensen2013}. 


\bibliography{limfast_paper1}{}
\bibliographystyle{aasjournal}

\end{document}